\documentclass[letterpaper,11pt]{article}
\pdfoutput=1
\usepackage{jheppub}
\usepackage{physics}
\usepackage{relsize}
\usepackage{tikz}
\usepackage{pgfplots}
\usepackage{comment}
\usepackage{amsmath}
\usepackage{amssymb}
\usepackage{xspace}
\usepackage{cancel}
\usepackage{tcolorbox}
\usepackage{hyperref}
\makeatletter
\def\@fpheader{\relax}
\makeatother

\usepackage[utf8]{inputenc}

\newcommand{\bigdot}{\raisebox{-0.5ex}{\scalebox{2}{$\cdot$}}}

\definecolor{blue3}{RGB}{31,119,180}
\definecolor{red3}{RGB}{214,39,40}
\definecolor{orange3}{RGB}{255,127,14}
\definecolor{green3}{RGB}{44,160,44}

\def\bfk{\textbf{k}}
\def\bfe{\textbf{e}}
\def\bfx{\textbf{x}}

\def\P{$\bf{P}$\xspace}

\def\RR{$\bf{RR}$\xspace}
\def\CRT{$\bf{CRT}$\xspace}

\newcommand{\C}{$\bf{C}$\xspace}

\numberwithin{equation}{section}

\newcommand*\diff{\mathop{}\!\mathrm{d}}

\allowdisplaybreaks[1]
\title{No-go Theorem for Cosmological Parity Violation}

\author{Ayngaran Thavanesan}
\affiliation{Department of Applied  Mathematics and Theoretical Physics, University of Cambridge,\\Wilberforce Road, Cambridge, CB3 0WA, United Kingdom.}
\affiliation{Kavli Institute for Theoretical Physics, Santa Barbara, CA 93106, USA.}
\emailAdd{at735@cantab.ac.uk}
\date{March 2024}
\abstract{A no-go theorem for parity-violation in even $D$-dimensional spacetimes invariant under $ISO(d)$ and dilatations (as well as the implications for odd $D$) is derived. For the case of real massless scalar and gravitons (as well as any massless even integer spin-$s$ field) at $\mathcal{I}^+$, the reality of wavefunction coefficients in Fourier space to all orders in perturbation theory (any order in loops) coming from a scale-invariant, IR- and UV-finite theory, which start from the initial \CRT-invariant Bunch-Davies state in the infinite past, is proven. From this, it is inferred that a parity-odd correlator with any massless scalar fields and even integer spin-$s$ fields vanishes in the presence of any number of interactions of massless fields. The same is true for correlators with an even number of conformally-coupled and massless odd integer spin-$s$ external fields, which is used to derive the cosmological analogue of Furry's theorem. The fundamental implications of \CRT symmetry for theories with chemical potentials, such as Chern-Simons and Axion inflation, are also discussed. Given the recent interest in observational claims of parity-violation detection, these results provide clear constraints on parity-violating models of inflation and establish the measurement of any parity-odd correlator as an exceptionally sensitive probe of new physics beyond vanilla inflation.

\vspace{1.4cm}
{\begin{center}
\textit{This work is dedicated to the memory of Lt. Colonel Santosham Master (Umainesan Kanpathippillai) and Kuganesan Kanpathippillai.}
\end{center}}}
 
\begin{document}

\maketitle

\section{Introduction}
In theoretical cosmology, we often look towards distant regions of the universe to investigate high-energy physics, searching for signatures of how the universe behaved in its earliest moments. One key aspect is the potential violation of parity (the symmetry between left-handed and right-handed configurations) in the primordial universe. Given that the late universe's evolution is governed by general relativity which respects parity symmetry, detecting its violation could provide valuable insights into the universe's initial conditions which are governed by fundamental physics.

This paper seeks to probe parity violation by examining various cosmological correlators --- quantities that describe statistical relationships between different points in the universe. Correlators involving fields with spin, like gravitational waves (described by tensor fields), can show evidence of parity violation through simple signals like two-point functions. Specifically, different helicities of gravitons can have varying strengths in their corresponding signals. Higher-order correlators, such as the bispectrum (three-point function) and trispectrum (four-point function), are also sensitive to parity violation. However, in the simpler case of scalar fields, parity-violating signals only begin to emerge in four-point correlations due to symmetry restrictions.

This paper delves into the correlation functions of primordial scalar and tensor (general integer spin-$s$) fields and develops several no-go theorems --- results that show under what conditions such parity-violating signals are forbidden. This matches and generalises previous more restricted no-go theorems in the literature~\cite{Liu:2019fag,Cabass:2022rhr,Stefanyszyn:2023qov} obtained from either explicit computations or by using symmetries and the perturbative unitarity constraints of the Cosmological Optical Theorem~\cite{COT,Goodhew:2021oqg,Cespedes:2020xqq}. The paper also briefly discusses examples where these conditions are relaxed, allowing parity violation to emerge in the scalar trispectrum and higher-order correlation functions, or the graviton correlation functions.

The motivation for this work is not purely theoretical: recent experimental results hint at possible signs of parity violation in polarisation data of the Cosmic Microwave Background (CMB)~\cite{Minami:2020odp,Orlando:2022rih,Philcox:2023ypl} as well as the large-scale structures like galaxies~\cite{Cahn:2021ltp,Hou:2022wfj,Philcox:2022hkh,Philcox:2023uor}. Though these findings are tentative~\cite{Krolewski:2024paz}, they highlight the importance of understanding how parity violation could manifest in the scalar and tensor sectors of cosmology~\cite{Niu:2022fki,Creque-Sarbinowski:2023wmb,Garcia-Saenz:2023zue,Coulton:2023oug,Zhang:2023scq,Jazayeri:2023kji,Fujita:2023inz,Akama:2024bav,Inomata:2024ald,Moretti:2024fzb,Reinhard:2024evr}. By using insights from the recently derived Cosmological CPT Theorem~\cite{Goodhew:2024eup}, this research clarifies when parity-odd correlators can and cannot arise and thus motivate concrete examples of physical mechanisms that can produce these signals.

The main result of this paper is the following No-go Theorem for Cosmological Parity Violation. In even $D=d+1$-spacetime dimensions, parity-odd correlators at the $\eta=0^-$ boundary of inflation \emph{cannot} be generated at \emph{tree or loop-level} by models with:
\begin{itemize}
    \item \emph{only} massless scalar fields and even-spin fields (e.g. gravitons) satisfying the Bunch-Davies vacuum (can also have even number of massless odd-spin fields, e.g. photons, or conformally coupled fields in external legs);
    \item locally \CRT-invariant Lagrangians;
    \item IR-finite and UV-finite boundary wavefunction coefficients $\psi_{n}$.
\end{itemize}
If the last two criteria are met, then a parity-odd correlator $B_n^{\rm{PO}}$ at the $\eta=0^-$ boundary of inflation can only be sourced by the \emph{factorised} contribution from internal massive, spinning fields (see~\cite{Stefanyszyn:2023qov,Stefanyszyn:2024msm} for a more detailed discussion). This result also has a striking parallel with Furry’s theorem in quantum electrodynamics, which forbids correlators involving an odd number of photons due to charge conjugation symmetry~\cite{Furry:1937zz}. In the cosmological setting, a similar result holds: a similar principle applies: if the external legs contain an odd number of odd-spin fields (such as photons), and the theory respects both charge conjugation and \CRT symmetry, then the corresponding wavefunction coefficients are purely imaginary, implying the vanishing of the parity-even part of the correlator. This constitutes a cosmological analogue of Furry’s theorem in de Sitter space, introducing new structural constraints on the space of allowed observables.

At \emph{loop-level}, the no-go theorem can be circumvented in the UV-divergent case due to the terms proportional to an $i\pi$ correction coming from logarithmic UV-divergences. In dimensional regularisation (dim-reg), the dimension $d$ is analytically continued to non-integer dimensions, and the logarithmic UV-divergence turns into a $1/\delta$ divergence (for IR-finite $1$-loop diagrams). If one adopts the regularisation scheme of~\cite{Senatore:2009cf}, where the mass of the field is not renormalised, the phase of the wavefunction coefficients $\psi_{n}$ are real, and thus $B_n^{\rm{PO}}$ still vanishes. However, if one uses a regularisation scheme, where the mass of the field is renormalised\footnote{In this approach, the mass of the field is also renormalised to keep the order of the Hankel function $\nu$ fixed, which ensures that the integrals can be computed analytically in the dim-reg parameter $\delta$. The alternative prescription proposed in~\cite{Senatore:2009cf} does not renormalise the mass of the field, resulting in the order of the Hankel function for the case of massless fields in $d=3+\delta$-dimensions becoming $\nu=d/2=(3+\delta)/2$. However, this prevents the integrals from being computed analytically. Consequently, the authors carry out an expansion in $\delta$ before computing the integrals which introduces non-trivial corrections at various points in the calculation. This makes the overall answer non-analytic in $\delta$ and makes a direct comparison with the phase formula in \eqref{eqn:PhaseFormula} and \eqref{eqn:PhaseArgument} nebulous.}, the phase of $\psi_{n}$ becomes complex, which when expanded can cancel the $1/\delta$ divergence, using the simple fact that any generic complex number $A$ can be expressed as $A=|A|e^{i \arg(A)}$. For IR-finite $1$-loop diagrams we thus find
\begin{equation}
    \lim_{\delta \to 0} \left[ |\psi^{(L=1)}|e^{i\arg(\psi^{(L=1)})} \right] \sim \frac{1}{\delta} (1 \pm i\pi \delta + O(\delta^2)) = \frac{1}{\delta} \pm i\pi + O(\delta) \, .
\end{equation}
This phenomenon was found in~\cite{Lee:2023jby}, where the authors computed a parity-odd contribution to the scalar trispectrum using the in-in formalism. However, this is a generic feature of UV-divergent loop diagrams and will be explored further in~\cite{Thavanesan:2025tha}. It is still an open question whether the two regularisation schemes are consistent with one another, but here we conjecture that they are equivalent up to counterterms, and remain agnostic whether the resulting parity-odd correlator is physical. It should also be noted that UV-finite loop diagrams such as those coming from a one-site loop explored in~\cite{Lee:2023jby}, will not lead to a finite imaginary piece to $\psi_n$ which could source a parity-odd correlator.

On the other hand in odd $D=d+1$-spacetime dimensions, parity-even correlators at the $\eta=0^-$ boundary of inflation \emph{cannot} be generated at \emph{tree level or even-loop level} by models with
\begin{itemize}
    \item \emph{only} massless scalar fields and even-spin fields (e.g. gravitons) satisfying the Bunch-Davies vacuum (can also have even number of massless odd-spin fields, e.g. photons, or conformally coupled fields in external legs);
    \item locally \CRT-invariant Lagrangians;
    \item IR-finite and UV-finite boundary wavefunction coefficients $\psi_{n}$.
\end{itemize}
An interesting observation is that, in  $D=d+1$-spacetime dimensions, $\psi_n \in i\mathbb{R}$ at tree-level (and at any even-loop level) for pure gravity, massless scalars, and any massless even spin- s  fields. However, normalisability requires $\psi_n \in \mathbb{R} < 0$. This suggests the intriguing conclusion that bulk loop effects are essential for constructing a consistent theory involving massless scalars and gravitons.

More broadly, this work generalises previous no-go theorems~\cite{Liu:2019fag,Cabass:2022rhr,Stefanyszyn:2023qov}, extending their applicability to any spacetime dimension and any massive integer spin-$s$  field, as long as the conformal dimension $\Delta \in \mathbb{R}$.

A key takeaway is that parity violation is tightly constrained in scale-invariant theories with standard cosmological vacua. However, relaxing these constraints—such as introducing time-dependent couplings or non-standard vacua (e.g. the ghost condensate scenario~\cite{GhostCondensate,Arkani-Hamed:2003juy,Cabass:2022rhr,Cabass:2022oap}) - enables the possibility of parity-odd signals. Similarly, interactions involving massive spinning fields or breaking exact scale invariance can generate non-vanishing parity-odd signals.

The results presented in this paper provide a theoretical framework for understanding parity violation in the early universe, offering guidance for future observational efforts to detect such signals. These findings could unveil new physics beyond the standard models of cosmology and particle physics.

\subsection*{Plan of Paper}
The structure of the paper is as follows:

We begin in Section \ref{sec:WFU} by reviewing the cosmological wavefunction (also referred to as the wavefunction of the universe in the literature) and its corresponding wavefunction coefficients. We discuss how these coefficients are computed for various types of theories, identifying the wavefunction for both scalar and integer spin fields, and defining the corresponding wavefunction coefficients at the boundary at the end of inflation. In this section, we also introduce the wavefunction coefficients, leaving their symmetry transformations to be explored in later sections.

To lay the groundwork for subsequent results, in Section \ref{sec:CosmoCorrelators} we distinguish between wavefunction coefficients and cosmological correlators—two concepts that do not entirely overlap (only cosmological correlators are the relevant observables in cosmology). We demonstrate that parity-odd correlators are sourced by the imaginary part of the wavefunction coefficient.

In Section \ref{sec:WFUConstraints}, we review the Cosmological CPT Theorem and examine how discrete symmetries in cosmology constrain the wavefunction coefficients. In particular, we show that the \CRT is sufficient to determine a formula for the phase of the wavefunction coefficients at future infinity. 

Using this phase formula, we identify precisely when the wavefunction acquires an imaginary part, leading to a no-go theorem for parity violation in cosmology in even spacetime dimensions, which we present in Section \ref{sec:NoGoParityViolation}. We also formulate a cosmological analogue of Furry’s theorem and discuss the distinct behaviour that arises in odd spacetime dimensions.

Finally, we conclude in Section \ref{sec:Summary} with a discussion of the implications of this no-go theorem for cosmology and inflation. We also explore potential ways to circumvent the no-go theorem and outline possible directions for future work.

\section{The Wavefunction of the Universe}\label{sec:WFU}
In this section, we will provide a review of how the cosmological wavefunction (also referred to as the wavefunction of the universe in the literature), and its corresponding wavefunction coefficients, are computed for various types of theories. In this paper, we will be discussing theories with interactions involving scalars and integer-spin fields and will be working with real fields without loss of generality since any complex field can be written in a basis of real fields. As explained in~\cite{Goodhew:2024eup} it should be possible to extend the analysis to half-integer-spin fields (see e.g.~\cite{Sun:2021thf,Pethybridge:2021rwf,Schaub:2023scu,Schaub:2024rnl,Letsios:2020twa,Letsios:2022tsq,Letsios:2023awz,Letsios:2023qzq,Letsios:2023tuc,Liu:2019fag,Sou:2021juh,Tong:2023krn,Chowdhury:2024snc,Baumann:2024ttn} for discussions of spinors in de Sitter space and cosmology) but we leave this as a direction for future work.

\subsection{Scalars}\label{eqn:Scalars}
The action of a massive scalar in an arbitrary $D=d+1$-dimensional spacetime background is 
\begin{align}
    S_{\sigma} = \int \diff^D x \sqrt{-g} \, \mathcal{L} &= \int \diff^D x \sqrt{-g} \left(\mathcal{L}_{\text{free}} + \mathcal{L}_{\text{int}} \right) \\ 
    &= \int \diff^D x \sqrt{-g} \left( -\frac{1}{2} g^{\mu \nu} \partial_{\mu}\sigma \partial_{\nu}\sigma -\frac{1}{2}m^2\sigma^2 + \mathcal{L}_{\text{int}} \right) \, ,
\end{align}
where the interaction terms in the Lagrangian density have been grouped into $\mathcal{L}_{\text{int}}$ and separated from the free part $\mathcal{L}_{\text{free}}$. The background metric for a spatially flat FLRW spacetime in the expanding patch is given by
\begin{equation}\label{eqn:FLRWmetric}
    \diff s^2 = - \diff t^2 + a^2(t)\diff{\bf x}^2 \,\, ,
\end{equation}
A perfect fluid has an equation of state
\begin{equation}\label{eqn:EOS}
    p=w\rho \, ,
\end{equation}
where $p$ is the pressure, $\rho$ is the mass density of the fluid in the comoving frame, $w$ is the equation of state parameter, with $w=-1$ for a cosmological constant dominated universe, i.e. de Sitter. When we solve the Friedmann equations for the case of a cosmological constant-dominated universe, we find the scale factor to be
\begin{equation}\label{eqn:dSscalefactor}
    a(t) \propto e^{Ht} \, ,
\end{equation}
where $H$ is the Hubble parameter related to the de Sitter length $\ell$ and the cosmological constant $\Lambda$ in the usual way 
\begin{equation}\label{eqn:dSHubble}
    H^2=\frac{1}{\ell^2}=\frac{2\Lambda}{d(d-1)} \, ,
\end{equation}
and the cosmological time $t \in (-\infty,\infty)$. Let us now take the background metric to be exact de Sitter and work in the Poincaré (inflationary) patch, for which the metric is given by
\begin{equation}\label{eqn:dSPoincarémetric}
    \diff s^2 = - \diff t^2 + a^2(t)\diff{\bf x}^2\,\,, \qquad a(t) = e^{H t} \,\, ,
\end{equation}
and this metric describes one half of the global geometry of de Sitter space using planar spacelike slices growing from the infinite past at $t=-\infty$ to the boundary of de Sitter space $\mathcal{I}^+$ (or equivalently the end of inflation) at $t=\infty$. One can redefine these co-ordinates $\diff t = a(\eta)\diff \eta$ to rewrite \eqref{eqn:dSPoincarémetric} as a conformally flat metric, 
\begin{equation}\label{eqn:dSPoincaréconformalmetric}
    \diff s^2 = a^2(\eta)({- \diff\eta^2} + \diff{\bf x}^2)\,\,, \qquad a(\eta) = -\frac{1}{H \eta} \,\, ,
\end{equation}
since computing the late-time wavefunction is more convenient to do so in conformal time $\eta=-e^{-Ht} \in (-\infty,0^-)$, where the perturbations evolve from the far past at $\eta=-\infty$ to $\mathcal{I}^+$ at $\eta=0$. The free part of the action for a massive scalar for this particular background metric \eqref{eqn:dSPoincaréconformalmetric} is thus given by
\begin{equation}
    S_{\sigma, \, \text{free}} = \int \diff \eta \diff^d {\bf x} \, \left[a^{d-1}(\eta) \right] \left(\frac{1}{2}(\sigma')^2 - \frac{1}{2}c_{s}^2 \partial_{i}\sigma \partial_{i}\sigma - \frac{1}{2}m^2\sigma^2  \right) \, ,
\end{equation}
where primes denote derivatives with respect to conformal time $\eta$ and we have allowed for an arbitrary, constant speed of sound $c_{s}$ which signals the fact we are allowing for dS boosts to be spontaneously or explicitly broken.\footnote{When the speed of sound differs from the speed of light appearing in the metric, $  c_{s}\neq 1 $, the sound cone is not invariant under de Sitter boosts, a fact which can be simply seen in the flat-space limit, where de Sitter boosts reduce to Lorentz boosts.}

The equation of motion for a free massive scalar field in de Sitter is thus given by
\begin{equation} \label{eqn:EOMFreeField}
    \sigma''-\frac{d-1}{\eta}\sigma'+c_{s}^2\nabla^2\sigma+\frac{m^2}{H^2\eta^2}\sigma = 0
\end{equation}
The equation of motion in the late-time limit $\eta \to 0^-$ simplifies to 
\begin{equation}\label{eqn:EOMdSMassiveFreeField}
    \sigma'' -\frac{(d-1)}{\eta}\sigma' \; \cancelto{0}{+ c_{s}^2\nabla^2 \sigma} \; + \frac{m^2}{H^2\eta^2}\sigma = 0 \, ,
\end{equation}
and has simple power-law solutions
\begin{align}
    \lim_{\eta \rightarrow 0^-}\sigma(\eta,\textbf{x})&=\overline{\sigma}_{+}(\textbf{x})\eta^{\Delta^+}+\overline{\sigma}_{-}(\textbf{x})\eta^{\Delta^-} \, , \label{eqn:dSLateTimeBulkScalarField} \\
    &=\overline{\sigma}_{+}(\textbf{x})\eta^{\Delta}+\overline{\sigma}_{-}(\textbf{x})\eta^{d-\Delta} \, , \label{eqn:dSLateTimeBulkScalarField2}
\end{align}
where $\Delta^+$ and $\Delta^-$ obey the standard quadratic Casimir relation for massive scalar representations of the Euclidean conformal group $SO(d+1,1)$
\begin{equation}\label{eqn:dSCasimirScalar}
    m^2/H^2=m^2\ell^2=\Delta(d-\Delta) \, ,
\end{equation}
with the conformal dimension defined in the usual way $\Delta = d/2 + \nu$, $\nu=\sqrt{d^2/4 - m^2/H^2} \,$ by defining $\overline{\sigma}_{+}$ with $\Delta^+ \equiv \Delta$ and $\overline{\sigma}_{-}$ (with $\Delta^- \equiv d-\Delta$).

Working in momentum space, one can write the quantum free field operator as 
\begin{equation} \label{eqn:FreeField}
    \hat{\sigma}({\bf k}, \eta) = \sigma^{-}(k, \eta) a_\bfk + \sigma^{+}(k,\eta) a_{-\bfk}^\dagger \, ,
\end{equation}
where the mode functions $\sigma^{\pm} (k, \eta)$ correspond to solutions of the free classical equation of motion and are given by
\begin{equation}\label{eqn:ModeFunctionsMassive}
    \sigma^+(k,\eta)=i\,\frac{\sqrt{\pi} H}{2} e^{-i\frac{\pi}{2}(\nu+\frac{1}{2})}\, \left( \frac{ -\eta}{c_{s}}  \right)^{\frac{d}{2}} H^{(2)}_{\nu}(-c_{s}  k\eta) \, , \qquad \sigma^-(k,\eta)=(\sigma^+(k,\eta))^* \, ,
\end{equation}
where $\nu=\sqrt{d^2/4-m^2/H^2}$ denotes the order of the Hankel function, and by simply setting $m=0$, one can find the mode functions for massless fields to be
\begin{equation}\label{eqn:ModeFunctionsMassless}
    \phi^+(k,\eta)=i\,\frac{\sqrt{\pi} H}{2} e^{-i\frac{\pi}{2}(\frac{d+1}{2})}\, \left( \frac{ -\eta}{c_{s}}  \right)^{\frac{d}{2}} H^{(2)}_{d/2}(-c_{s}  k\eta) \, , \qquad \phi^-(k,\eta)=(\phi^+(k,\eta))^* \, .
\end{equation}
For the purposes of this work, we will impose the boundary condition that the fields started at the beginning of inflation in a non-excited initial state, specifically the Bunch-Davies vacuum. This corresponds to imposing a boundary condition on the fields, that they vanish in the infinite past ($\eta \to -\infty$)
\begin{equation} \label{eqn:BDFreeField}
    \lim_{\eta \to -\infty} \hat{\sigma}({\bf k}, \eta) = \sigma^{+}(k,\eta) a_{-\bfk}^\dagger \, .
\end{equation}
In reality this leads to attributing an $i\epsilon$-prescription to the magnitude of the spatial momenta, i.e. $k=\sqrt{k \cdot k}=\tilde{k}(1-i\epsilon) \in \mathbb{C}$, where $\tilde{k} \in \mathbb{R} \geq 0$.

\subsection{Spinning Fields}\label{sec:SpinningFields}
The mode functions for graviton fluctuations take the same form as \eqref{eqn:ModeFunctionsMassless} (with $c_{s}=1$) with the addition of polarisation tensors $e^{h}_{ij}({\bf k})$, with $h = \pm 2$, as required by little group scaling. This is because, for each polarisation mode, the equation of motion is that of a massless scalar. The polarisation tensors satisfy the following conditions:
\begin{align}\label{pol1}
    e_{ii}^{h}({\bf k}) - k^{i}e^{h}_{ij}({\bf k})&=0 & \text{(transverse and traceless)} \, , \\
    e_{ij}^{h}({\bf k}) - e_{ji}^{h}({\bf k}) &=0 & \text{(symmetric)} \, , \\
    e_{ij}^{h}({\bf k})e_{jk}^{h}({\bf k}) &=0 & \text{(lightlike)} \, , \\
    e^{h}_{ij}({\bf k})e^{h'}_{ij}({\bf k})^{\ast} - 4\delta_{hh'} &=0 & \text{(normalisation)} \, , \label{eqn:normeps}  \\
    e_{ij}^{h}({\bf k})^{\ast} - e_{ij}^{h}(-{\bf k}) &=0 &\text{($  \gamma_{ij}(x) $ is real)}  \, .\label{eqn:poln}
\end{align} 
For generic spin-$s$ fields, it is convenient to use the following notations and conventions. In $D=d+1$-dimensional spacetime, for traceless\footnote{To consider fields with a non-zero trace, one can subtract the trace and treat it as an additional scalar field.}, integer spin-$s$ fields we use the free action developed in~\cite{Bordin:2018pca} and discussed in the context of the cosmological bootstrap in~\cite{Goodhew:2021oqg,Goodhew:2022ayb,Goodhew:2024eup}:
\begin{equation}\label{eqn:SpinningFreeAction}
    S=\int \diff^D x \, \left[a(\eta)\right]^{d-1}\frac{1}{2s!}\left[\left(\sigma_{i_1\dots i_s}' \right)^2-c_s^2\left(\partial_j\sigma_{i_1\dots i_s}\right)^2-\delta c_s^2 \left(\partial^j\sigma_{ji_2\dots i_s}\right)^2-m^2a^2\left(\sigma_{i_1\dots i_s}\right)^2\right] \, .
\end{equation}
The totally-symmetric, traceless tensor $\sigma_{i_1 \dots i_s}$ has spatial indices $i_1=1,\ldots,d$ which span the $d$-dimensional spacelike hypersurface orthogonal to the $\eta$ coordinate\footnote{The Effective Field Theory of Inflation (EFToI)~\cite{Cheung:2007st} is derived by considering a
theory which is only invariant under spatial diffeomorphisms, for which there is a preference for the co-ordinate choice used in unitary gauge, where the time coordinate is chosen to coincide with the surfaces of constant value of the field $\sigma_{i_1 \dots i_s}$. The EFToI thus encapsulates a generic class of models of inflation where spatial diffeomorphisms are preserved. \eqref{eqn:SpinningFreeAction} can be written in a covariant way by using the Goldstone boson $\pi$ of time translations to upgrade the spatial tensor $ \sigma_{i_1 \dots i_s} $ to a covariant spacetime tensor. The coupling of $ \sigma_{i_1 \dots i_s} $ to $\pi$ is also dictated by this constructions but we will not need this here.}. $ \sigma_{i_1 \dots i_s} $ has $(2s+1)$ components, which each create states (“particles") with helicities $ 0,\pm 1, \dots, \pm s$, and we have enforced invariance under dilatations by including inverse factors of the scale factor for each coordinate derivative. Following~\cite{Goodhew:2021oqg,Goodhew:2022ayb} we Fourier transform and diagonalise this using the helicity modes, $\sigma_h$, defined by:
\begin{equation}\label{eqn:PhiFourierTreansform}
    \sigma_{i_1 \ldots i_s}(\eta; x)=\int_{\bfk} e^{i \bfk \cdot \bfx} \sum_{h=-S}^{S} \bfe_{i_1 \ldots i_s}^{h}(\bfk) \sigma_{h}(\eta; \bfk) \, ,
\end{equation}
These helicity tensors are defined as an outer product of helicity vectors,
\begin{equation}
    \bfe^h_{i_1\dots i_s}=\bfe^{h_1}_{i_1}\dots \bfe^{h_s}_{i_s} \,,
\end{equation}
which satisfy the following relations:
\begin{align}
    \bfe^{h}_{i}(\bfk) \left[\bfe^{h'}_{i}(\bfk)\right]^{*} - 4\delta_{hh'} &=0 & \text{(orthogonality and normalisation)}\,, \label{eqn:PolOrtho}
    \\
    \left[\bfe_{i}^h(\bfk)\right]^{*}-\bfe_{i}^h(-\bfk) &= 0 & \text{($  \sigma_{i_1 \ldots i_s}(x) $ is real)}  \, . \label{eqn:PolReal}
\end{align} 
Note that these fields are not assumed to be transverse, $h$ is allowed to take $d$ different values including $0$ where $\bfe^0$ is proportional to the momentum. The contributions from the other helicity modes are therefore transverse by the orthogonality condition.

The equation of motion for a massive spin-$s$ $\sigma_{i_1,\ldots,i_s}$ field in exact de Sitter is found to be
\begin{equation}
    \sigma_{i_1 \ldots i_s}''-\frac{d-1}{\eta}\sigma_{i_1 \ldots i_s}'+\left(\frac{m^2/H^2-(s-2)(s+d-2)}{\eta^2}+c_{s}^2\nabla^2 \right)\sigma_{i_1 \ldots i_s} = 0 \, .
\end{equation}
The equation of motion for a free massless spin-$s$ in the late-time limit $\eta \to 0$ simplifies to 
\begin{equation}\label{eqn:EOMFLRWMassiveFreeSpinningField}
    \sigma_{i_1 \ldots i_s}'' -\frac{d-1}{\eta}\sigma_{i_1 \ldots i_s}' + \left(\frac{m^2/H^2-(s-2)(s+d-2)}{\eta^2}\right)\sigma_{i_1 \ldots i_s} \; \cancelto{0}{+ c_{s}^2\nabla^2 \sigma_{i_1 \ldots i_s}} \; = 0 \, ,
\end{equation}
and has simple power-law solutions
\begin{align}
    \lim_{\eta \rightarrow 0^-}\sigma_{i_1 \ldots i_s}(\eta,\textbf{x})&=\overline{\sigma}_{i_1 \ldots i_s,+}(\textbf{x})\eta^{\Delta^+}+\overline{\sigma}_{i_1 \ldots i_s,-}(\textbf{x})\eta^{\Delta^-} \, , \label{eqn:dSLateTimeBulkSpinningField} \\
    &=\overline{\sigma}_{i_1 \ldots i_s,+}(\textbf{x})\eta^{\Delta}+\overline{\sigma}_{i_1 \ldots i_s,-}(\textbf{x})\eta^{d-\Delta} \, , \label{eqn:dSLateTimeBulkSpinningField2}
\end{align}
where $\Delta^+$ and $\Delta^-$ obey the standard quadratic Casimir relation for massive integer spin-$s$ representations of the Euclidean conformal group $SO(d+1,1)$
\begin{equation}
    m^2/H^2=m^2\ell^2=(\Delta+s-2)(d+s-2-\Delta) \, ,
\end{equation}
with the conformal dimension defined in the usual way $\Delta = d/2 + \mu$, $\mu=\sqrt{(d+2s-4)^2/4 - m^2/H^2} \,$ by defining $\overline{\sigma}_{i_1 \ldots i_s,+}$ with $\Delta^+ \equiv \Delta$ and $\overline{\sigma}_{i_1 \ldots i_s,-}$ (with $\Delta^- \equiv d-\Delta$). 

We parameterise the wavefunction, $\Psi$, at conformal time $\eta_0$ in terms of the helicities of the integer spin field as
\begin{align}\label{eqn:WavefunctionOfTheUniverse}
    \Psi[\eta_0;\sigma({\bf k})] = \text{exp}\left[-\sum_{n=2}^{\infty} \frac{1}{n!} \sum_{h_i=\pm} \int_{\bfk_{1}, \ldots, \bfk_{n}} \psi^{h_{1} \ldots h_{n}}_{n}(\eta_0;{\bfk}) (2\pi)^d \delta^d \left(\sum \bfk_a \right) \sigma_{h_1}(\bfk_{1}) \ldots \sigma_{h_n}({\bfk}_{n}) \right] \, .
\end{align}
Spatial translations and spatial rotations ensure that wavefunction coefficients can be written as a product of a \textit{helicity factor}, which is an $SO(d)$ invariant function of helicity vectors and spatial momenta, multiplied by a \textit{trimmed wavefunction coefficient}~\cite{Cabass:2021fnw,Cabass:2022jda} (see also~\cite{Maldacena:2011nz} where the authors introduced a spinor helicity formalism to describe gravitational waves with circular polarisation in cosmology) which is only a function of the magnitudes of the momenta in the literature:
\begin{equation} \label{eqn:GeneralWFC}
    \psi^{h_{1} \ldots h_{n}}_{n} = \text{(tensor structure)} \times \text{(trimmed wavefunction coefficient)} \, .
\end{equation}
We take all coefficients appearing in the tensor structure to be real and therefore include any factors of $i$ that might appear when converting to momentum space, or simply as part of the Feynman rules, in the trimmed part which we will denote as $\psi_{n}$ for brevity. For a general $\psi^{h_{1} \ldots h_{n}}_{n}$, we have 
\begin{equation}\label{GeneralPsin}
    \psi^{h_{1} \ldots h_{n}}_{n} = \left[\bfe^{h_1}(\bfk_1) \ldots \bfe^{h_n}(\bfk_n) \, \bfk_1^{\alpha_1} \ldots \bfk_n^{\alpha_n}\right] \psi_n,
\end{equation}
for some integer $\alpha_i$. Note that we can choose for \emph{both parity-even and parity-odd} tensor structures to be invariant with respect to the discrete symmetries of \CRT, \RR and $\mathbf{D}$ which we will be discussing in this paper.\footnote{Had we used another convention where the tensor structures scaled in a non-trivial way with Dilatations $\mathbf{D}$, then the scalar component of the field would have to scale inversely, in order to leave the field $\sigma_{i_1 \ldots i_s}$ invariant.} Hence, all our results extend directly from the scalar $\psi_n$ to the tensor case.

\subsection{The Boundary Wavefunction}\label{sec:BoundaryWFU}
We are interested in scenarios where dS boosts are broken since it is known that these symmetries could not have been exact in the early universe, and large non-Gaussianities are associated with a large breaking of boosts~\cite{Green:2020ebl}. We will take the remaining symmetries of the dS group to be exact: spatial translations, spatial rotations and dilations. A general interaction vertex with $n$ fields, scalars and spinning fields, therefore takes the schematic form
\begin{align}
S_{\text{int}} =  \int \diff \eta \diff^d {\bf x} \, a(\eta)^{D - N_{\text{deriv}}} \partial^{N_{\text{deriv}}} \varphi^{n}\,,
\end{align}
where $\varphi$ can be any spinning field, $\partial$ stands for either time derivatives $\partial_{\eta}$ or spatial derivatives $\partial_{i}$, and $N_{\text{deriv}}$ is the total number of derivatives. Spatial derivatives and the spinning fields' indices are contracted with the $SO(d)$ invariant objects $\delta_{ij}$ and $\epsilon_{ijk}$ and the overall number of scale factors is dictated by scale invariance. Here and throughout this and following sections we use $\varphi(\bfk)$ to schematically denote scalars and integer spin-$s$ fields, with $SO(d)$ indices suppressed, and each of these fields satisfies $\varphi(\bfk) = \varphi(-\bfk)^{\ast}$ which follows directly from \eqref{eqn:FreeField}, \eqref{eqn:ModeFunctionsMassive} and \eqref{eqn:poln}.

\begin{figure}\label{fig:psi5}
    \centering
\tikzset{every picture/.style={line width=0.75pt}} 

\begin{tikzpicture}[x=0.75pt,y=0.75pt,yscale=-1,xscale=1]
\draw    (300,60) -- (725,60) ;
\draw    (375,175) -- (650,175) ;
\draw    (350,60) -- (375,175) ;
\draw    (400,60) -- (375,175) ;
\draw    (512.5,60) -- (512.5,175) ;
\draw    (625,60) -- (650,175) ;
\draw    (675,60) -- (650,175) ;
\draw [shift={(350,60)}] [color={rgb, 255:red, 0; green, 0; blue, 0 }  ][fill={rgb, 255:red, 0; green, 0; blue, 0 }  ][line width=0.75]      (0, 0) circle [x radius= 3.35, y radius= 3.35]   ;
\draw [shift={(400,60)}] [color={rgb, 255:red, 0; green, 0; blue, 0 }  ][fill={rgb, 255:red, 0; green, 0; blue, 0 }  ][line width=0.75]      (0, 0) circle [x radius= 3.35, y radius= 3.35]   ;
\draw [shift={(512.5,60)}] [color={rgb, 255:red, 0; green, 0; blue, 0 }  ][fill={rgb, 255:red, 0; green, 0; blue, 0 }  ][line width=0.75]      (0, 0) circle [x radius= 3.35, y radius= 3.35]   ;
\draw [shift={(625,60)}] [color={rgb, 255:red, 0; green, 0; blue, 0 }  ][fill={rgb, 255:red, 0; green, 0; blue, 0 }  ][line width=0.75]      (0, 0) circle [x radius= 3.35, y radius= 3.35]   ;
\draw [shift={(675,60)}] [color={rgb, 255:red, 0; green, 0; blue, 0 }  ][fill={rgb, 255:red, 0; green, 0; blue, 0 }  ][line width=0.75]      (0, 0) circle [x radius= 3.35, y radius= 3.35]   ;
\draw (250,60) node [anchor=center]{$\eta _{0} =0$};
\draw (350,60) node [anchor=south]{$k_1$};
\draw (400,60) node [anchor=south]{$k_2$};
\draw (512.5,60) node [anchor=south]{$k_3$};
\draw (625,60) node [anchor=south]{$k_4$};
\draw (675,60) node [anchor=south]{$k_5$};
\draw (375,175) node [anchor=north]{$g(\varphi ')^3$};
\draw (512.5,175) node [anchor=north]{$g(\varphi ')^3$};
\draw (650,175) node [anchor=north]{$g(\varphi ')^3$};

\end{tikzpicture}

\caption{Tree-level exchange Feynman diagram for $\psi_5$ generated by a $(\varphi ')^3$ interaction.}
\end{figure}
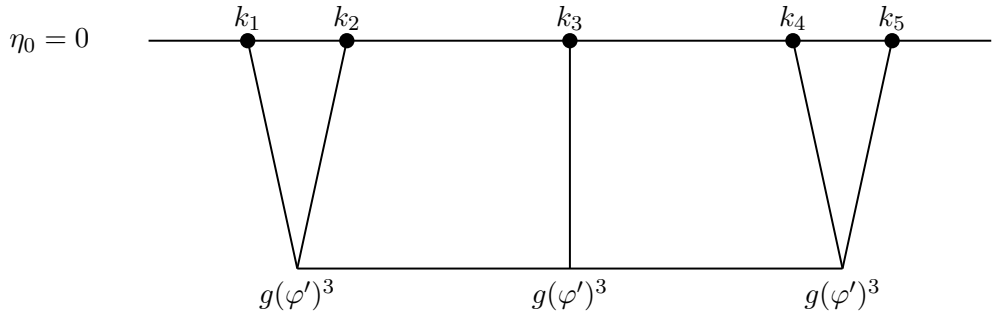

Let us start with defining the late-time wavefunction which is one of the objects of interest in cosmology, wherein the weak coupling approximation can be parameterised in terms of a series expansion
\begin{align}
    \Psi[{\eta_0;\varphi}]&=\int_{\phi(-\infty)=0}^{\phi(\eta_0)=\varphi} \mathrm{D}\phi \, e^{iS[\phi]} \\
    &=\exp{-\sum_{n=2}^\infty \left[\prod_{a=1}^n \int \frac{\diff^d k_a}{(2\pi)^d}\varphi (\bfk_a) \right] \psi_{n}(\eta_0;\bfk)} \, , \label{eqn:ParameterisedWFU}
\end{align}
where $\{ k \}$ collectively denotes the external energies~\footnote{The literature refers to the magnitude of a spatial momentum vector as ``energy'' despite the absence of time translation symmetry in cosmology, since in cosmological amplitudes/observables they play the analogous role to energy in flat-space amplitudes.} $k_a=|\bfk_a|$, $\{ \bfk \}$ collectively denotes their spatial momenta, and $\varphi(\bfk)$ collectively represents all fields in the theory with indices suppressed. The wavefunction coefficients are also dependent on their internal energies, which are a function of the external lines spatial momenta as required by momentum conservation; a wavefunction coefficient $\psi_n$ with $n$ external can have an arbitrary number of $m < n$ internal fields. For example, a tree-level $\psi_5$ coming from the exchange diagram depicted in Figure \ref{fig:psi5} with 2 internal lines and a cubic $(\varphi ')^3$ interaction and coupling $g$ (where $\varphi$ can be a scalar or spinning field with any mass) in each vertex, has the form, 
\begin{equation}\label{eqn:psi5}
    \psi_5 \sim g^3 e^{i\arg(\psi_5)} \frac{\text{Poly}_{d+(a+b+c+d)-\alpha}(k_1,k_2,k_3,k_4,k_5)}{(k_T)^a(k_1+k_2+|\bfk_1+\bfk_2|)^b(k_3+|\bfk_1+\bfk_2|+|\bfk_4+\bfk_5|)^c(k_3+k_4+|\bfk_4+\bfk_5|)^d} \, ,
\end{equation}
with $a$ being the order of the leading total-energy $k_T$-pole\footnote{The total-energy is defined as the sum of all the external energies in any given Feynman-Witten diagram. For our example in \eqref{eqn:psi5} the total energy is given by $k_T=k_1+k_2+k_3+k_4+k_5$.}, $b,c,d$ being the order of the partial-energy poles, and $d+(a+b+c+d)-\alpha$ the degree of the polynomial which is fixed by scale invariance. We emphasise this point as we will be utilising the analyticity of both the external $k_a=|\bfk_a|$ and internal energies $|\sum \bfk_i|\equiv\sqrt{\sum \bfk_i}$~\footnote{These are sometimes denoted as $s\equiv|\sum \bfk_i|\equiv\sqrt{\sum \bfk_i}$ in the literature.}, where the internal line connects two vertices and $\bfk_i$ denotes all the external momenta entering one vertex.

Notice that this parameterisation does not require any saddle-point approximation of the bulk path integral that defines $\Psi$. In fact, the wavefunction coefficients can be found non-perturbatively from
\begin{equation}
    \begin{split}
        \psi_n(\eta_0;\bfk) \equiv \psi'_n(\eta_0;\bfk) (2 \pi)^d \delta^d \bigg(\sum_{a=1}^n \bfk_a \bigg)&=- \frac{\delta^n \log\Psi [\eta_0;\varphi]}{ \delta \varphi_{\bfk_1}\cdots \delta \varphi_{\bfk_n}  } \bigg|_{\varphi  = 0} \, ,
    \end{split}
\end{equation}
where the prime in $\psi'_n$ is used to explicitly highlight that these wavefunction coefficients contain the momentum-conserving $\delta$-function\footnote{This is a result of translation invariance, which in position space, implies that correlation functions are invariant under spatial translations, i.e.,
\begin{equation}
    \langle \mathcal{O}_1(\bfx_1 + \mathbf{a}) \cdots \mathcal{O}_n(\bfx_n + \mathbf{a}) \rangle = \langle \mathcal{O}_1(\bfx_1) \cdots \mathcal{O}_n(\bfx_n) \rangle
\end{equation}
for any constant vector $\mathbf{a}$. Now, if we consider the Fourier transform of an $n$-point function
\begin{equation}
    \langle \mathcal{O}_1(\mathbf{k}_1) \cdots \mathcal{O}_n(\mathbf{k}_n) \rangle = \int \prod_{a=1}^n \diff^d x_a \, e^{-i \mathbf{k}_a \cdot \bfx_a} \langle \mathcal{O}_1(\bfx_1) \cdots \mathcal{O}_n(\bfx_n) \rangle \, ,
\end{equation}
and perform a simultaneous shift of all positions $\bfx_a \to \bfx_a +\mathbf{a}$. Translation invariance ensures that the correlator is unchanged, while the plane waves pick up a phase:
\begin{equation}
    \prod_{a=1}^n e^{-i \mathbf{k}_a \cdot (\bfx_a + \mathbf{a})} = e^{-i \mathbf{a} \cdot \sum_a \mathbf{k}_a} \prod_{a=1}^n e^{-i \mathbf{k}_a \cdot \bfx_a} \, .
\end{equation}
Since the position-space correlator is invariant, the only change is this overall phase. But this implies that the momentum-space correlator satisfies
\begin{equation}
    \langle \mathcal{O}_1(\mathbf{k}_1) \cdots \mathcal{O}_n(\mathbf{k}_n) \rangle = e^{-i \mathbf{a} \cdot \sum_a \mathbf{k}_a} \langle \mathcal{O}_1(\mathbf{k}_1) \cdots \mathcal{O}_n(\mathbf{k}_n) \rangle \, ,
\end{equation}
which can only hold for all $\mathbf{a}$ if
\begin{equation}
    \sum_{a=1}^n \mathbf{k}_a = 0 \, .
\end{equation}

Therefore, translation invariance implies that \emph{momentum is conserved}, and this is encoded as a delta function enforcing momentum conservation
\begin{equation}
    \langle \mathcal{O}_1(\mathbf{k}_1) \cdots \mathcal{O}_n(\mathbf{k}_n) \rangle = (2\pi)^d \delta_{\rm D}^{(d)}\left(\sum_{a=1}^n \mathbf{k}_a\right) \times \langle \mathcal{O}_1(\mathbf{k}_1) \cdots \mathcal{O}_n(\mathbf{k}_n) \rangle' \, ,
\end{equation}
where the reduced correlator $\langle \mathcal{O}_1(\mathbf{k}_1) \cdots \mathcal{O}_n(\mathbf{k}_n) \rangle'$ contains the dynamical information not fixed by translation invariance.} whilst those without the $\delta$-function will be denoted by $\psi_n$. Upon renormalisation, $\psi_n$ can be computed to any desired order in perturbation theory, including any number of loops. In this work, we focus on the natural observables of the Poincar\'e patch of de Sitter and of inflationary cosmology, namely correlation functions of the equal-time product of fields at the future conformal boundary $\eta_0 \to 0$.

In the context of dS/CFT, the wavefunction is treated as a partition function $\Psi[g_{ij},\varphi,\ldots]=Z[g_{ij},\varphi,\ldots]$ and satisfies the standard rules of a generating functional, where the wavefunction coefficients are the dS/CFT correlators, e.g. for the 2-point stress tensor correlator we have
\begin{equation}
    \langle T_{ij}(\bfk_1) T_{lm}(\bfk_2) \rangle_{\eta_0} = \frac{\delta^2 \log\Psi [\eta_0;g_{ij}]}{ \delta g_{ij}(\bfk_1)\delta g_{lm}(\bfk_2)  } \bigg|_{g_{ij} = 0} = -\psi'_2(\eta_0;\bfk) (2 \pi)^d \delta^d (\bfk_1+\bfk_2) \,\,.
\end{equation}
We insert here a brief aside about scaling dimensions as it is important for the arguments which follow. By considering the wavefunction of the universe at the future boundary of dS, where the time dependence often trivialises, we can derive direct constraints on the wavefunction coefficients. This future boundary is where the reheating surface is understood to live in the inflationary paradigm. Therefore, such constraints are of particular cosmological relevance.

In the late-time limit, the bulk fields take the form
\begin{align}
    \lim_{\eta \rightarrow 0^-}\varphi(\eta,\textbf{x})&=\overline{\varphi}_{+}(\textbf{x})\eta^{\Delta^+}+\overline{\varphi}_{-}(\textbf{x})\eta^{\Delta^-} \, , \label{eqn:dSLateTimeBulkField} \\
    &=\overline{\varphi}_{+}(\textbf{x})\eta^{\Delta}+\overline{\varphi}_{-}(\textbf{x})\eta^{d-\Delta} \, , \label{eqn:dSLateTimeBulkField2}
\end{align}
where for the case of scalars, the mass $m^2/H^2=m^2\ell^2=\Delta(d-\Delta)$ is related to the conformal dimension in the usual way $\Delta = d/2 + \nu$, $\nu=\sqrt{d^2/4 - m^2/H^2} \, $\footnote{Note for AdS this relation is $m^2 \ell_{AdS}^2=\Delta(\Delta-d)$. In AdS $\Delta^{\pm} \in \mathbb{R}$ for scalar fields of any mass. In dS for scalar fields with mass $m^2/H^2 > d^2/4$, $\Delta^{\pm} \in \mathbb{C}$ which is known as the principal series; fields with mass $m^2/H^2 \leq d^2/4$, $\Delta^{\pm} \in \mathbb{R}$ which is known as the complementary series.}. For massive spin-$s$ fields, the mass and spin are related to the conformal dimension by $\Delta = d/2 + \mu$, $\mu=\sqrt{(d+2s-4)^2/4 - m^2/H^2} \, $. We define $\overline{\varphi}_{+}$ with $\Delta^+ \equiv \Delta$ and $\overline{\varphi}_{-}$ (with $\Delta^-=d-\Delta$). For heavy scalar fields ($m>dH/2$) and heavy spin-$s$ fields ($m>(d+2s-4)H/2$), $\Delta$ is complex and so these boundary operators do not represent a self-adjoint basis\footnote{Of course, we could have made the choice to expand the fields in a self-adjoint way but this would sacrifice boundary conformal invariance as such an expansion would necessarily mix terms of different weights. See e.g. \cite{Anous:2020nxu,Anninos:2023lin,Joung:2006gj,Sengor:2019mbz,Sengor:2021zlc,Sengor:2022hfx,Sengor:2022lyv,Sengor:2022kji,Sengor:2023buj,Dey:2024zjx} for discussions regarding the principal series.}. We will therefore restrict our discussion to light fields and leave a more comprehensive examination, including heavy fields, to future work. In this case, $\Delta$ is real and positive so the $\overline{\varphi}_{-}$ term dominates. Starting from the expansion of the wavefunction in terms of its coefficients, \eqref{eqn:ParameterisedWFU}, we can similarly define a ``boundary" wavefunction\footnote{Although this is referred to as a ``boundary" wavefunction in the literature, this is not really a wavefunction living at the $\mathcal{I}^+$ where $\eta=0^-$, since from \eqref{eqn:dSLateTimeBulkField} it is clear that only massless fields would survive in the limit of $\eta=0^-$. The more accurate description is that the coefficients of the wavefunction have had their $\eta$-dependence stripped off.} which is a functional of these boundary fields $\overline{\varphi}_{-}$ and has no explicit $\eta$-dependence
\begin{equation}\label{eqn:ParameterisedBoundaryWFU}
    \overline{\Psi}[\overline{\varphi}_{-}(\bfk_a)] = \exp\left\{-\sum_{n=2}^\infty \int \left[\prod_{a=1}^n\frac{\diff^dk_a}{(2\pi)^d}\overline{\varphi}_{-}(\bfk_a) \right] \overline{\psi}_n(\bfk) \delta^d\left(\sum \bfk_a\right)\right\} \, , 
\end{equation}
where we have introduced boundary wavefunction coefficients\footnote{These can be interpreted as the correlators of the operators of some conjectured CFT living at the boundary, i.e. dS/CFT.} denoted by $\overline{\psi}_n$ to distinguish them from the bulk wavefunction coefficients $\psi_n$.\footnote{To obtain the boundary wavefunction coefficient we differentiate the boundary wavefunction $\overline \Psi$ with respect to the sources $\overline{\varphi}_{-}$ which have dimension $d - \Delta$
\begin{equation}
    \overline{\psi}_n(\bfk) \equiv \lim_{\overline{\varphi}_{-}\to 0} \frac{\delta^n }{\delta^n \overline{\varphi}_{-}} \log \overline{\Psi}[\overline{\varphi}_{-}(\bfk_a)] \, .
\end{equation}
}
The choice to Taylor expand the wavefunction in the field operator with dimension $d-\Delta$ rather than using the seemingly more sensible choice of labelling this operator's dimension $\Delta$ is purely conventional. It is standard to refer to $\Delta^+$ as \emph{the dimension} of a field in cosmology (for example a massless field has dimension $\Delta=d$) but it turns out that this is not the part of the field that survives on the boundary.

Let us now count the scalings of each term in the exponent one by one to see how $\overline{\psi}_n(\bfk_a)$ scales under $\bfk \to \lambda \bfk$ which in turn will scale the internal and external energies $k \to \lambda k$. The $d$-dimensional $\diff^dk$ measures each scale with volume and thus they will scale with an overall factor of $\lambda^{nd}$. This cancels with an equivalent scaling of the Fourier transformed $\overline{\varphi}_{-}(\bfk_a)$ which gives an overall scaling of $\lambda^{-nd}$.\footnote{This is in the same way that in \eqref{eqn:ParameterisedWFU} the original bulk field $\varphi_{\bfk_a}$ scaled as $\lambda^{d}$ resulting in the bulk wavefunction coefficient scaling inversely to the $\delta^d\left(\sum \bfk_a\right)$.} Each boundary field must contribute an extra scaling like $\eta^{\Delta-d}$ to ensure the Scale Invariance of the bulk field \eqref{eqn:dSLateTimeBulkField2}. The result of this is that in addition to the factor coming from the Fourier Transform the fields contribute a further $\lambda^{dn-\sum_\alpha \Delta_\alpha}$. Finally, the $\delta^d(\bfk_1+\ldots+\bfk_n)$ scales with inverse volume $\lambda^{-d}$.\footnote{Usually in the context of computations in de Sitter, the $\delta({\bfk_1}+\ldots+{\bfk_n})$-function is omitted but we remind the reader of its importance in terms of arguments involving Scale Invariance, as it scales with inverse volume, as well as its necessity for ensuring momentum conservation which comes from translation invariance in position space.} We know that the exponent of \eqref{eqn:ParameterisedBoundaryWFU} must be dimensionless, thus by dimensional analysis, any IR-finite boundary wavefunction coefficient will scale as
\begin{equation}\label{eqn:BoundarySI}
    \overline{\psi}_n(\lambda \bfk)=\lambda^{d} \lambda^{-nd} \lambda^{\sum_{\alpha} \Delta_{\alpha}}\overline{\psi}_n(\bfk) = \lambda^{d(1-n)+\sum_{\alpha} \Delta_{\alpha}}\overline{\psi}_n(\bfk) \, .
\end{equation}

To simplify notation we will drop the bar/overline notation and write $\overline{\psi}$ and $\overline{\phi}$ as $\psi$ and $\phi$ from this point onwards.

\section{Cosmological Correlators}\label{sec:CosmoCorrelators}
A powerful property of the wavefunction is that it contains all the information regarding the quantum processes that occur during inflation and thus fundamentally related to the observations we make on the night sky in cosmology with a specific dictionary translating information from the wavefunction to observations and vice versa. In particular, one can use the wavefunction to compute equal-time expectation values, i.e. cosmological correlators (also known as in-in correlators) using the usual quantum mechanics formula, i.e.
\begin{equation}
    B_{n}(\bfk) \equiv \langle \varphi(\bfk_1) \ldots \varphi(\bfk_n)  \rangle = \frac{\int \mathrm{D} \varphi ~ \Psi \Psi^{\ast} ~ \varphi(\bfk_1) \ldots \varphi(\bfk_n)}{\int \mathrm{D} \varphi ~ \Psi \Psi^{\ast}}\,,
\end{equation}

\noindent Here we have not made a distinction between the dependence of the wavefunction coefficients on the set of spatial momenta $\{\bfk\}$ and their norms $\{ k \}$, but in general we will work away from the physical configuration and treat $\{\bfk\}$ and $\{ k \}$ as independent objects, for reasons that will become clear. Since we will be making a purely non-perturbative statement which is true for all exchange and loop diagrams, it is also important to note that there is a possible dependence on internal energies (sometimes denoted as $\{s \}$ in the literature) which are related to external energies by momentum conservation, as demonstrated in \eqref{eqn:psi5}. We are going to use \CRT to constrain the form of the probability distribution $ \Psi \Psi^{\ast}$. Now from this perturbative expression for the wavefunction, we have
\begin{align}
    - \log(\Psi \Psi^{\ast}) &=  \left( \sum_{n=2}^{\infty} \frac{1}{n!} \int_{{\bfk}_{1}, \ldots, {\bfk}_{n}} \psi_{n}(\bfk)\varphi({\bfk}_{1}) \ldots \varphi({\bfk}_{n}) \right) \nonumber \\ 
    &+ \left( \sum_{n=2}^{\infty} \frac{1}{n!} \int_{{\bfk}_{1}, \ldots, {\bfk}_{n}} \psi_{n}(\bfk)\varphi({\bf k}_{1}) \ldots \varphi({\bfk}_{n}) \right)^{\ast} \\
    &= \left( \sum_{n=2}^{\infty} \frac{1}{n!} \int_{{\bfk}_{1}, \ldots, {\bfk}_{n}} \psi_{n}(\bfk)\varphi({\bfk}_{1}) \ldots \varphi({\bfk}_{n}) \right) \nonumber  \\
    &+ \left( \sum_{n=2}^{\infty} \frac{1}{n!} \int_{{\bfk}_{1}, \ldots, {\bfk}_{n}} \psi^{\ast}_{n}(\bfk)\varphi(-{\bfk}_{1}) \ldots \varphi(-{\bfk}_{n}) \right) \, .
\end{align}
If we  change the integration variables on the final line by sending $\{ \bfk \} \rightarrow  \{- \bfk \}$ we have 
\begin{equation}
    - \log(\Psi \Psi^{\ast}) = \sum_{n=2}^{\infty} \frac{1}{n!} \int_{{\bfk}_{1}, \ldots, {\bf{k}}_{n}} [\psi_{n}(\bfk)+ \psi^{\ast}_{n}(- \bfk)]\varphi({\bf k}_{1}) \ldots \varphi({\bf k}_{n}) \,.
\end{equation}
It follows from Gaussian integral formulae that the resulting correlators arising from Feynman diagrams in perturbation theory, are given by
\begin{equation}\label{WFtoCorrelator}
    B_{n}(\bfk) =  -\frac{\psi_n (\bfk) +\psi_n^\ast (-\bfk)}{\prod_{a=1}^{n} \psi_2 (\bfk) +\psi_2^\ast (-\bfk)} + \textit{factorised} \, ,
\end{equation}
where the lower order factorised terms are $0$ for the contact diagram contribution to $B_n$ and come in the same type of linear combination of $\psi_n$'s, e.g. for $B_4$ at tree-level we find
\begin{equation}
    B_4=-\frac{1}{\prod_{a=1}^4 \psi_2 (\bfk) +\psi_2^{\ast} (-\bfk )}\left[\psi_4 (\bfk) +\psi_4^\ast (-\bfk) - \frac{\left(\psi_3 (\bfk) +\psi_3^\ast (-\bfk) \right) \left(\psi_3 (\bfk) +\psi_3^\ast (-\bfk)\right)}{\psi_2 (\bfk) +\psi_2^\ast (-\bfk)} \right] \, .
\end{equation}
For parity-even interactions, we thus have
\begin{equation} \label{eqn:PEcorelator}
    B_{n}(\bfk) =  -\frac{\psi_n (\bfk) +\psi_n^\ast (\bfk)}{\prod_{a=1}^{n} \psi_2 (\bfk) +\psi_2^\ast (\bfk)} + \textit{factorised} \, ,
\end{equation}
the numerator and denominator are simply $2 \text{Re} ~ \psi'_{n}$ and $2 \text{Re} ~ \psi'_{2}$ respectively in which case our expression matches the one that usually appears in the literature. Let us now define a new variable\footnote{In recent literature~\cite{Cabass:2022rhr}, the connection between the wavefunction and the density matrix has been explored in terms of their coefficients, as they essentially correspond to the components of a density matrix. This connection becomes explicit in the basis of field eigenstates, which satisfy \(\hat{\varphi}(\bfx)\ket{\varphi} = \varphi(\bfx)\ket{\varphi}\). Here, we explicitly indicate the operator with a hat for this equation, although we will omit it in subsequent expressions for simplicity. Using such eigenstates, we can insert two resolutions of the identity to express the density matrix:
\begin{equation}
\rho = \int \mathrm{D}\varphi \mathrm{D}\bar{\varphi} , |\varphi\rangle \langle\varphi|\rho|\bar{\varphi}\rangle \langle\bar{\varphi}| = \int \mathrm{D}\varphi \mathrm{D}\bar{\varphi} , \rho_{\varphi\bar{\varphi}} , |\varphi\rangle \langle\bar{\varphi}|,.
\end{equation}
From this, it follows that the components  of the operator  in the field basis are given by:
\begin{equation} \label{eqn:DensityMatrixField}
    \rho_{\varphi\bar{\varphi}} = \langle\varphi|\rho|\bar{\varphi}\rangle = \langle\varphi|\Omega\rangle \langle\Omega|\bar{\varphi}\rangle = \Psi[\varphi] \Psi[\bar{\varphi}]^{\ast} \, ,
\end{equation}
where  is the field-theoretic wavefunction corresponding to the state \(\ket{\Omega}\). Here, we remind the reader that we are working in a basis of real fields, where the reality condition implies that, in momentum space, \(\varphi(\bfk) = \varphi^{\ast}(-\bfk)\).} related to the wavefunction coefficients
\begin{align}
    \rho_n (\bfk) &= \psi_n (\bfk) +\psi_n^\ast (- \bfk)\,\,.
\end{align}
In perturbation theory, correlators can be computed in terms of the $\rho_n$'s. For example, at tree level we have
\begin{align} \label{WFUtoCorrelators}
    B_2\equiv P&=-\frac{1}{\rho_2}\,\,, &
    B_4&=-\frac{1}{\prod_{a=1}^4 \rho_2(\bfk_a)} \left[\rho_4 - \frac{\rho_3 \rho_3}{\rho_2} \right]\,\,, &    B_n^{\text{contact}}&=-\frac{\rho_n}{\prod_{a=1}^n \rho_2(\bfk_a)} \,\, ,
\end{align}
where $P$ is the power spectrum. However, we are interested in parity-violating correlators, i.e. the parity-odd contribution to the correlator, which is related to wavefunction coefficients via the density matrix coefficients $\rho_n$. For the parity-even contribution to the correlator we have
\begin{align}
    \rho_n^{\rm{PE}} (\bfk) &= \frac{1}{2} \left[ \rho_n (\bfk) + \rho_n (- \bfk)\right] \\
    &= \frac{1}{2} \left[ \psi_n (\bfk) + \psi_n^\ast (\bfk) + \psi_n (- \bfk) + \psi_n^\ast (- \bfk) \right] \, ,
\end{align}
where $\rho_n^{\rm{PE}}(-\bfk)=\rho_n^{\rm{PE}}(\bfk)$. Thus we see that $\rho_n^{\rm{PE}}$ must be purely real to find a non-vanishing parity-even correlator. For the parity-odd contribution to the correlator we have
\begin{align}
    \rho_n^{\rm{PO}} (\bfk) &= \frac{1}{2} \left[ \rho_n (\bfk) -\rho_n (- \bfk)\right] \\
    &= \frac{1}{2} \left[ \psi_n (\bfk) -\psi_n^\ast (\bfk) + \psi_n^\ast (- \bfk) - \psi_n (- \bfk) \right] \label{eqn:ParityOddCorrelator}\, ,
\end{align}
where $\rho_n^{\rm{PO}}(-\bfk)=-\rho_n^{\rm{PO}}(\bfk)$. Thus we see that $\rho_n^{\rm{PO}}$ must be purely imaginary if we are to find a non-vanishing parity-odd correlator:
\begin{align}
    \rho_n^{\rm{PO}} &\in i \mathbb{R} & \rho_n^{\rm{PE}} &\in  \mathbb{R}\,\,.
\end{align}
From this we can infer that $ \psi_{n} \in \mathbb{R} \implies \rho_n^{\rm{PO}}=0$. Our no-go theorems will thus be based on asking when $\psi_{n}$, can be imaginary, specifically that if we find $\psi_{n} \in \mathbb{R}$ we can infer that the parity-odd cosmological correlator vanishes. We can see from \eqref{eqn:ParityOddCorrelator} that although an imaginary part to $\psi_n$ is necessary in order to have a non-zero $\rho_n^{\rm{PO}}$, if $\psi_n$ is purely parity-even, i.e. $\psi_n (-\bfk)=\psi_n (\bfk)$, we find
\begin{align}
    \rho_n^{\rm{PO}} (\bfk) &= \frac{1}{2} \left[ \psi_n (\bfk) -\psi_n^\ast (\bfk) + \psi_n^\ast (- \bfk) - \psi_n (- \bfk) \right] \\
    &= \frac{1}{2} \left[ \psi_n (\bfk) -\psi_n^\ast (\bfk) + \psi_n^\ast (\bfk) - \psi_n (\bfk) \right] \\
    &=0 \, .
\end{align} 
Hence an imaginary part to $\psi_n$ is \emph{necessary but not sufficient} to ensure a parity-odd correlator, since the imaginary part to $\psi_n$ must itself be parity-odd, i.e. $\psi_n (-\bfk)=-\psi_n (\bfk)$, by being sourced by a parity-odd interaction.

\section{Constraints on the Wavefunction}\label{sec:WFUConstraints}
One particular application of the Cosmological CPT Theorem~\cite{Goodhew:2024eup} is the constraint of discrete transformations at a \textit{single} asymptotic future boundary $\eta=0^-$, where the wavefunction coefficients have no dependence on time, i.e. the dS/CFT correlators. Let us look at each transformation individually.

A constraint corresponding to bulk unitary time evolution known as Reflection Reality \RR was derived in~\cite{Goodhew:2024eup}. \RR does not care about the time at which the wavefunction is evaluated, i.e. it is the same for both boundary and bulk wavefunction coefficients, and when expressed perturbatively:
\begin{equation}\label{eqn:BoundaryCOT} 
    \left[ \psi^{(L)}_n(\bfk) \right]^* = e^{i\pi [(d+1)L-1]} \psi^{(L)}_n(e^{-i\pi}\bfk) \, ,
\end{equation}
where the rotation $e^{-i\pi}\bfk$ not only rotates the spatial momenta $\{\bfk\}$, but also rotates their norms $\{ k \}$, since under this rotation the norm or ``energy'' transforms as
\begin{equation}
    k\rightarrow \sqrt{\bfk\cdot\bfk\,e^{-2i\theta}}=\sqrt{\bfk\cdot\bfk}\sqrt{e^{-2i\theta}} \, ,
\end{equation}
where the final line follows as $k$ is real. To ensure the single-valuedness of $k$ during this rotation we need to continuously follow a Riemann sheet of the square root (i.e. we will not add any factors of $2\pi$ to the exponent). Similarly the internal energies $\{s \}$ which are related to external energies by momentum conservation which exchange and loop diagrams will generically depend will transform as
\begin{equation}
    s\rightarrow \sqrt{(\bfk_1+\ldots+\bfk_n)\cdot(\bfk_1+\ldots+\bfk_n)\,e^{-2i\theta}}=\sqrt{(\bfk_1+\ldots+\bfk_n)\cdot(\bfk_1+\ldots+\bfk_n)}\sqrt{e^{-2i\theta}} \, .
\end{equation}

As was derived in \eqref{eqn:BoundarySI}, Scale Invariance of the boundary theory at $\mathcal{I}^+$ where $\eta=0^-$ tells us that the wavefunction coefficients must scale as
\begin{equation}\label{eqn:BoundarySI2}
   \psi^{(L)}_n(\lambda \bfk) = \lambda^{d(1-n)+\sum_{\alpha} \Delta_{\alpha}} \psi^{(L)}_n(\bfk) \, ,
\end{equation}
where $\psi_n^{(L)}$ denotes a perturbative wavefunction coefficient computed at some loop order $L$ (e.g. $L=0$ corresponds to tree-level and $L=1$ corresponds to $1$-loop order), i.e. it is the loop order of the bulk theory, and $\Delta_\alpha$ are the dimensions of the external fields. The Cosmological CPT theorem tells us that the $SO^+(1,1)$ boost which is the important continuous symmetry in the flat-space CPT theorem acts in the Poincaré patch as dilatations \textbf{D}, i.e. Scale Invariance $\lambda \in \mathbb{R}^+$, which we can analytically continue to $SO(2)$, provided we have a Hamiltonian bounded from below. In~\cite{Goodhew:2024eup}, it was demonstrated that the three discrete symmetries \RR, $\mathbf{D}^{\pm}_{-1}$ and \CRT form a group structure, and thus two of the three discrete symmetries can be combined to obtain the third. Under analytic continuation of $\lambda \in \mathbb{R}^+$ through the complex plane $\lambda \in \mathbb{C}$,\footnote{Given that $\psi^{(L)}_n$ is only analytic in the lower half-place $\psi^{(L)}_n \in \mathbb{C}^{-i}$, we can see from the group structure discussed in~\cite{Goodhew:2024eup} we must act with $\mathbf{D}_{-1}^{+}$ after \RR in order to obtain \CRT; or alternatively act with \RR after $\mathbf{D}_{-1}^{-}$.} we can then access $\lambda=e^{-i\pi}$, to conclude that:
\begin{equation}\label{eqn:DiscreteSI}
   \psi^{(L)}_n(e^{-i\pi}\bfk) = (-1)^{d(1-n) + \sum_\alpha \Delta_\alpha} \psi^{(L)}_n(\bfk) \, ,
\end{equation}
where for cases with fractional $d$ and $\Delta$ we have
\begin{equation}\label{eqn:DiscreteSI2}
   \textbf{D}_{-1}^{\pm} : \psi^{(L)}_n(e^{\mp i\pi}\bfk) = e^{\mp i\pi [d(1-n) + \sum_\alpha \Delta_\alpha]} \psi^{(L)}_n( \bfk) \, .
\end{equation}
We need to take into account that $\psi^{(L)}_n$ is analytic through the lower half-plane $\mathbb{C}^{-i}$. Given that the \RR \eqref{eqn:BoundaryCOT} transformation rotates $\psi^{(L)}_n$ counter-clockwise, we must use the $\textbf{D}_{-1}^{+}$ in \eqref{eqn:DiscreteSI2} to land on
\begin{equation}\label{eqn:Phase}
    \textbf{D}_{-1}^{+} \bigdot  \mathbf{RR}: \left[ \psi^{(L)}_n( \bfk) \right]^* = e^{i\pi [(d+1)L-1-d(1-n)-\sum_{\alpha} \Delta_{\alpha}]} \psi^{(L)}_n( \bfk) \, .
\end{equation}
We can also derive the same expression independently using \CRT, which tells us directly that
\begin{equation}\label{eqn:CPTphase}
    \left[\psi^{(L)}_n(\bfk)\right]^* = e^{i\pi [(d+1)(L-1) + dn - \sum_\alpha \Delta_\alpha]} \psi^{(L)}_n(\bfk) \, ,
\end{equation}
where in order to obtain \eqref{eqn:CPTphase} we have had to continue $1/\eta \propto k \equiv |\bfk|$ in the lower-half plane $\mathbb{C}^{-i}$, and as previously explained a factor of $\eta^{\sum_{\alpha}d-\Delta_{\alpha}}$ arises from the fields scaling as $\eta^{\Delta-d} \propto \lambda^{d-\Delta}$ at the boundary. We have now identified the discrete symmetries which the boundary wavefunction coefficients $\psi^{(L)}_n$ satisfy:
\begin{eqnarray}
\textbf{CRT}:& \psi^{(L)}_n(\bfk) \!&=\,\, e^{-i\pi[(d+1)(L-1)+dn-\sum_{\alpha} \Delta_{\alpha}]} \, \left[\psi^{(L)}_n(\bfk) \right]^* \, , \label{eqn:BoundaryCPT} \\
\textbf{D}_{-1}^{\, \pm}:& \,\, \psi^{(L)}_n(\bfk) \!&=\,\, e^{\pm i\pi[d(1-n)+\sum_{\alpha} \Delta_{\alpha}]} \, \psi^{(L)}_n(e^{\mp i\pi} \bfk) \, ,\\
\textbf{RR}:& \, \psi^{(L)}_n(\bfk) \!&=\,\, e^{-i\pi[(d+1)L-1]} \, \left[\psi^{(L)}_n(e^{-i\pi}\bfk) \right]^* \, ,
\end{eqnarray}
where the last condition from \RR holds for boundary correlators in any flat FLRW spacetime with a future conformal boundary.
We can use \eqref{eqn:BoundaryCPT} to solve for the phase of $\psi^{(L)}_n$ directly, obtaining the following phase formula for the boundary wavefunction coefficients:
\begin{equation}\label{eqn:PhaseFormula}
    e^{i \arg (\psi^{(L)}_n)} \equiv \frac{\psi^{(L)}_n(\bfk)}{|\psi^{(L)}_n(\bfk)|} 
    = \pm \sqrt{ \frac{\psi^{(L)}_n(\bfk)}{\psi^{(L)*}_n(\bfk)}} = \pm \, (-i)^{(d+1)(L-1)+dn-\sum_{\alpha} \Delta_{\alpha}} \, ,
\end{equation}
where there is a $\pm$ out front because \CRT cannot determine the overall real sign, because obtaining the phase involves taking a square root.\footnote{Although, for the 2-point function of a field which is weakly coupled in the bulk, this sign is fixed by normalisability.} Hence, we obtain the result quoted in~\cite{Goodhew:2024eup}:
\begin{equation}\label{eqn:PhaseArgument}
    \arg(\psi^{(L)}_n)
    = -\frac{\pi}{2}\!\left((d+1)(L-1)+dn-\sum_{\alpha} \Delta_{\alpha}\right) + \pi \mathbb{N} \, .
\end{equation}
Remarkably \eqref{eqn:PhaseFormula} and \eqref{eqn:PhaseArgument} will hold for any $\psi^{(L)}_n$ computed in cosmology provided that:
\begin{itemize}
    \item spacetime is de Sitter (possibly with boost-breaking terms);
    \item the Lagrangian is locally \CRT-invariant;
    \item the amplitude is UV- and IR-finite;
    \item and involves fields in representations with integer spin and real $\Delta$ (no spinors or principal series);
    \item all fields satisfy the Bunch-Davies vacuum.
\end{itemize}
Crucially, the phase of $\psi^{(L)}_n$ has no dependence on the details of the bulk interactions, e.g.~derivative couplings will have the same phase, provided the Feynman-Witten diagrams have the same external legs with the same $\Delta$'s. Thus both contact diagrams with the same external legs and those with internal lines (exchange and loop diagrams) will have the same phase, provided the exchanged field is not heavy, i.e. it does not fall within the principal series (see~\cite{Stefanyszyn:2023qov} for a more detailed discussion). 

The steps leading up to this phase formula implicitly assumed that the amplitudes were IR and UV-finite. For a typical IR-divergent amplitude, there is a term like $C \log(-\eta)$ which, upon rotating $\eta$ from $0^-$ to $0^+$, becomes $C \log (-e^{-i\pi}\eta)$  This provides an extra $-i\pi C$ shift that adjusts the IR-finite piece of the amplitude, which does not conform to \eqref{eqn:PhaseArgument}. But, the leading order IR-divergence will continue to satisfy \eqref{eqn:PhaseArgument}. An example of such a calculation can be found in~\cite{Goodhew:2024eup}.

It is also important to recognise that the conformal dimensions $\Delta_{\alpha}$ appearing in \eqref{eqn:PhaseArgument} are not protected quantities. In general, they can receive radiative corrections and thus run with couplings, leading to anomalous dimensions. The phase relation \eqref{eqn:PhaseArgument} should therefore be interpreted as applying to the \emph{effective}, i.e. loop-corrected, conformal dimensions at the order of perturbation theory considered.

For example, in loop-corrected wavefunction coefficients, logarithmic scale dependence such as $\log(k)$ or $\log(k\eta)$ appears, signalling a running dimension. As discussed in~\cite{Senatore:2009cf,Melville:2021lst}, a renormalised two-point wavefunction coefficient may take the form
\begin{equation}\label{eqn:psi2_Running}
    \psi_2 = k^{2\Delta_0-d}\left[1 + 2g\ln(k) + \mathcal{O}(g^2) \right] = k^{2\Delta_0-d} e^{2g\ln(k)} = k^{2(\Delta_0+g)-d} \, ,
\end{equation}
where $g$ is a small coupling and $\Delta_0$ the tree-level dimension. The loop-corrected or “effective” conformal dimension is thus
\begin{equation}
    \Delta(g) = \Delta_0 + g + \mathcal{O}(g^2) \, .
\end{equation}
This running modifies the phase and must be taken into account when applying the phase rule. Similar arguments apply for other wavefunction coefficients with external fields acquiring anomalous dimensions. Nevertheless, in theories where the correlators are free from such logarithmic corrections --- e.g. when $\Delta_\alpha$ are fixed and no divergences are present --- the formula \eqref{eqn:PhaseArgument} remains valid and yields powerful constraints on the imaginary parts of cosmological wavefunction coefficients.

\section{No-go Theorem for Cosmological Parity Violation}\label{sec:NoGoParityViolation}
By examining \eqref{eqn:PhaseFormula} we find that for any $d=2\mathbb{Z}+1$ (where $\mathbb{Z}$ means integer), $\psi^{(L)}_n \in \mathbb{R}$ when $dn-\sum_{\alpha} \Delta_{\alpha} \in 2\mathbb{Z}$ (this result was derived at tree-level independently for scalars, photons and gravitons in~\cite{Stefanyszyn:2023qov,Stefanyszyn:2024msm}). Given that, we know
\begin{itemize}
    \item[-] $\Delta=d$ for massless scalars;
    \item[-] $\Delta=(d+1)/2$ for conformally coupled scalars;
    \item[-] $\Delta=d+s-2$ for massless spinning fields,
\end{itemize}
 we arrive at the following:
\begin{tcolorbox}[colback=green!5,colframe=green!40!black,title=No-go Theorem for Cosmological Parity Violation]
    In $D=d+1$-spacetime dimensions, where $d=2\mathbb{Z}+1$, parity-odd correlators at the $\eta=0^-$ boundary of inflation \textbf{cannot} be generated at \textbf{tree or loop-level} by models with:
    \begin{itemize}
        \item[$\bigstar$] \textbf{only} massless scalar fields and even-spin fields (e.g. gravitons) satisfying the Bunch-Davies vacuum (can also have even number of massless odd-spin fields, e.g. photons, or conformally coupled fields in external legs);
        \item[$\bigstar$] locally \CRT-invariant Lagrangians;
        \item[$\bigstar$] IR-finite and UV-finite $\psi_{n}$.
    \end{itemize}
\end{tcolorbox}
If these criteria are met, then a parity-odd correlator $B_n^{\text{PO}}$ can only be sourced by \emph{factorised} contributions from internal massive spinning fields (see~\cite{Stefanyszyn:2023qov,Stefanyszyn:2024msm} for a more detailed discussion).

\subsection{UV Divergences}
As discussed previously, UV-finite loop diagrams—such as the one-site loop analysed in~\cite{Lee:2023jby} do not generate a finite $i\pi$ phase correction to the \CRT and \RR constraints, and hence do not contribute to parity-odd correlators.\footnote{Natural candidates for other UV-finite diagrams in cosmology would be those that give UV-finite amplitudes in flat space (see e.g.~\cite{Lee:2023kno}). The author thanks Lorenzo di Pietro for helpful discussions regarding this point.} At \emph{loop-level}, we generally encounter UV divergences, requiring us to work in non-integer  $d+\delta$ spatial dimensions (where $d \in \mathbb{Z}$ and $\delta$ is non-integer), which can allow for a non-vanishing $B_n^{\rm{PO}}$. 

\paragraph{Massless dim-reg}
In~\cite{Senatore:2009cf}, the authors do not renormalise the mass of the field at the outset. In this case, for massless fields in non-integer $d+\delta$ spatial dimensions, the order of the Hankel function becomes $\nu = (d+\delta)/2$, so that
\begin{equation}
    \Delta = \frac{d+\delta}{2} + \nu = \frac{d+\delta}{2} + \frac{d+\delta}{2} = d+\delta \, .
\end{equation}
In the massless dim-reg scheme, since the conformal dimension equals the number of spatial dimensions ($\Delta = d + \delta$), the phase of the one-loop wavefunction coefficient $\psi^{(L=1)}_n$ becomes
\begin{equation}\label{eqn:DimReg2PhaseArgument}
    \arg(\psi^{(L=1)}_n) = \pi \mathbb{N} \, .
\end{equation} 
Hence, no $i\pi$ phase arises because the wavefunction coefficients stay real. It should also be noted that the loop integrals cannot be computed exactly; instead, $\delta$ is expanded before integration, leading to different (non-analytic) behaviour.

In $d$ spatial dimensions, the wavefunction coefficients of scalar fields scale as\footnote{This follows from dimensional analysis of the Fourier transform of the position-space correlator, which scales like $\langle \mathcal{O}_1(x_1) \cdots \mathcal{O}_n(x_n) \rangle \sim \lambda^{-\sum_{\alpha} \Delta_{\alpha}}$ under dilatations $x_i \to \lambda x_i$. Fixing one point (say $x_n = 0$), the Fourier transform gives a factor of $\lambda^{d(n-1)}$ from the $d^{d(n-1)}x_i$ integration, leading to the overall momentum scaling. For example, the 2-point function $\langle \mathcal{O}_1(x)\mathcal{O}_2(0) \rangle \sim 1/|x|^{2\Delta}$ Fourier transforms to $\langle \mathcal{O}_1(\bfk)\mathcal{O}_2(-\bfk) \sim k^{2\Delta - d} \delta^d\left(\sum_{a=1}^n \textbf{k}_a\right)$.}
\begin{equation}\label{eqn:psin_tree}
    \left\langle \mathcal{O}_1(\bfk_1) \cdots \mathcal{O}_n(\bfk_n) \right\rangle ' \equiv \psi_n' \equiv \psi_n \, \delta^d\left(\sum_{a=1}^n \textbf{k}_a\right) = k^{d(1-n)+\sum_{\alpha} \Delta_{\alpha}} \, \delta^d\left(\sum_{a=1}^n \textbf{k}_a\right) \, ,
\end{equation}
From \eqref{eqn:psin_tree}, we see that in the massless dim-reg scheme in $d+\delta$ spatial dimensions, $\psi_n$ scales as
\begin{equation}\label{eqn:psin_masslessdimreg}
    \psi_n = k^{(d+\delta)(1-n)+n(d+\delta)} = k^{(d+\delta)} \, ,
\end{equation}
In this context, one can understand the appearance of $\log(k)$ terms in loop-corrected correlators as a signal of anomalous dimension running since
\begin{equation}\label{eqn:psin_masslessdimreg_running}
    \psi_n = k^{(d+\delta)} = k^d k^{\delta} = k^d e^{\delta\ln(k)} = k^d \left[1+\delta\ln(k)+\mathcal{O}(\delta^2) \right] \, ,
\end{equation}
For instance, the one-loop correction to the two-point function found in~\cite{Senatore:2009cf} has the form
\begin{equation}
    \psi_2 = k^{2\Delta_0 - d} \left[ 1 + 2g \log(k) + \mathcal{O}(g^2) \right] = k^{2(\Delta_0 + g) - d} + \dots \, ,
\end{equation}
which can be interpreted as a radiative shift in the conformal dimension,
\begin{equation}
    \Delta(g) = \Delta_0 + g + \mathcal{O}(g^2) \, ,
\end{equation}
as in standard QFT. A familiar example is the renormalisation of the electron mass in QED, where the physical (pole) mass receives quantum corrections from photon loops. In schemes such as dimensional regularisation with minimal subtraction, the running of the mass parameter with respect to the renormalisation scale $\mu$ is governed by the anomalous dimension of the mass. At one-loop, this results in a scale-dependent bare mass:
\begin{equation}
    m(\mu) = m_0 \left[1 + \frac{\alpha}{\pi} \ln\left(\frac{\mu}{m_0}\right) + \mathcal{O}(\alpha^2) \right] \, ,
\end{equation}
where $m_0$ is the renormalised mass at a reference scale, and $\alpha=e^2/4\pi \approx 1/137$ is the fine-structure constant, which is the loop expansion parameter in QED. Although the physical pole mass remains invariant under changes in $\mu$, this $\ln(\mu/m_0)$ dependence captures how the mass parameter must vary with scale to keep physical observables fixed. This scale dependence is conceptually analogous to the appearance of $\ln(k)$ in cosmological correlators, which reflects the running of operator dimensions due to quantum corrections.

\paragraph{Massive dim-reg}
An alternative prescription, used in the calculations of~\cite{Melville:2019tdc,Lee:2023jby}, the authors used non-integer dimensions for dimensional regularisation (dim-reg) following the prescription detailed in Appendix C of~\cite{Melville:2019tdc}. In this scheme, which we will refer to as \textbf{massive dim-reg}, the field mass is renormalised to keep the order of the Hankel function $\nu$ fixed, where for scalar (and spin $s=2$) fields we find
\begin{equation}
    \nu = \frac{d}{2} = \sqrt{\frac{(d+\delta)^2}{4}-\frac{m^2}{H^2}} \implies m = H\sqrt{\frac{\delta(2d+\delta)}{4}} \, ,
\end{equation}
allowing the integrals to be computed analytically in the dim-reg parameter $\delta$. Note that in this choice of regualrisation scheme, working in $d-\delta$ spatial dimensions will result in $m = i H\sqrt{\delta(2d+\delta)/4} \implies m^2 = -H^2\delta(2d+\delta)/4<0$, i.e. a tachyonic mass. Hence in the massive dim-reg regularisation scheme one must be careful to analytically continue the spatial dimensions within the region $d+\delta \geq d$, in order to preserve bulk unitarity. Consequently, the conformal dimension for massless scalar fields becomes
\begin{equation}
    \Delta = \frac{d + \delta}{2} + \nu = \frac{d + \delta}{2} + \frac{d}{2} = \frac{2d + \delta}{2} = d + \frac{\delta}{2} \, .
\end{equation}
As a result, the phase of the 1-loop wavefunction coefficient $\psi^{(L=1)}_n$ becomes
\begin{equation}\label{eqn:DimReg1PhaseArgument}
    \arg(\psi^{(L=1)}_n) = \frac{n\pi \delta}{4} + \pi \mathbb{N} \, .
\end{equation}
In dim-reg, the logarithmic UV divergence manifests as a $1/\delta$ pole (assuming the loop diagram is IR-finite), which contributes to the imaginary part of the phase upon expansion of the exponential. Specifically, for any complex number $A = |A| e^{i \arg(A)}$, we find
\begin{equation}
    \lim_{\delta \to 0} \left[ |\overline{\psi}^{(L=1)}| e^{i \arg(\overline{\psi}^{(L=1)})} \right] \sim \frac{1}{\delta} (1 + i\pi \delta + \mathcal{O}(\delta^2)) = \frac{1}{\delta} + i\pi + \mathcal{O}(\delta) \, .
\end{equation}
A manifestation of this generic mechanism for a particular UV-divergent loop diagram was identified in~\cite{Lee:2023jby}, where the authors computed a parity-odd contribution to the scalar trispectrum using the in-in formalism. 

From \eqref{eqn:psin_tree}, we see that in the massive dim-reg scheme in $d+\delta$ spatial dimensions, $\psi_n$ scales as
\begin{equation}\label{eqn:psin_massivedimreg}
    \psi_n = k^{(d+\delta)(1-n)+nd+\frac{n\delta}{2}} = k^{d+\delta-\frac{n\delta}{2}} \, ,
\end{equation}
Hence, the absence of a $\ln(k)$ term for $\psi_2$ in massive dim-reg is simply explained by the fact that
\begin{equation}\label{eqn:psi2_no_logk}
    \psi_2 = k^{d+\delta-\frac{2\delta}{2}} = k^{d} \, ,
\end{equation}
where $d \in \mathbb{Z}$ and thus has no expansion in the exponent, i.e. a running of dimensions (anomalous dimensions), which can give rise to a $\ln(k)$ term. Further implications of this result will be explored in~\cite{Thavanesan:2025tha}.

\paragraph{Comparison of massless and massive dim-reg}
The running of dimension \eqref{eqn:psin_masslessdimreg_running} found in  the massless dim-reg scheme is completely analogous to the effective shift in $\Delta$ introduced by the massive dim-reg prescription (with fixed $\nu = 3/2$), where
\begin{equation}
    \Delta = \frac{d}{2} + \nu = \frac{3 + \delta}{2} + \frac{3}{2} = \frac{6 + \delta}{2} = 3 + \frac{\delta}{2} \, .
\end{equation}
On the other hand, the massless dim-reg prescription, where the mass is not adjusted, yields
\begin{equation}
    \Delta = \frac{d}{2} + \nu = \frac{3 + \delta}{2} + \frac{3 + \delta}{2} = 3 + \delta \, ,
\end{equation}
which corresponds to the presence of a running dimension without absorbing it into the Hankel index $\nu$.

Both schemes ultimately capture the same physical effect: the running of $\Delta$ with respect to the coupling or the regulator. What differs is how this running manifests --- in the massless scheme, through explicit logarithmic scale dependence; in the massive scheme, through implicit shifts in the conformal dimension via $\nu$. Reconciling these regularisation schemes and clarifying their implications for cosmological parity violation will be explored further in~\cite{Thavanesan:2025tha}.

Overall, resummation of loops can further modify the effective mass and thus the conformal dimension, leading to additional corrections that lie outside the assumptions of our parity no-go theorem:
\begin{equation}
    \Delta(g) = \Delta_0 + \mathcal{O}(g) \, .
\end{equation}
This reinforces the importance of understanding anomalous dimensions and effective IR behaviour when evaluating the robustness of symmetry-based constraints in cosmology. However, it is important to note that not all fields experience such running. In particular, massless gauge bosons and gravitons are protected by their underlying gauge and diffeomorphism symmetries, respectively~\cite{Ward:1950xp,Takahashi:1957xn,Deser:1969wk,Slavnov:1972fg,Taylor:1971ff,tHooft:1974toh,Birrell:1982ix,PeskinSch,Weinberg:1996kr}. In standard QFT, gauge invariance forbids the generation of a mass term for the photon, ensuring that the corresponding conserved current remains exactly conserved, and that the conformal dimension of its dual boundary operator remains fixed at $\Delta=d-1$ under renormalisation. Similarly, diffeomorphism invariance protects the graviton from acquiring a mass, preserving the conservation and tracelessness of the stress-energy tensor, and thereby fixing its conformal dimension at $\Delta=d$ under renormalisation. These symmetries prevent radiative corrections from altering the tensorial structure or scaling behaviour of the kinetic terms.

By contrast, scalar fields generically acquire corrections to their mass under renormalisation unless protected by an additional symmetry. For example, some inflationary models impose an (approximate) shift symmetry $\phi \to \phi+c$ to suppress quantum corrections to the inflaton potential and ensure radiative stability of the scalar mass\footnote{The author thanks Harry Goodhew for helpful discussions on the shift symmetries of the inflaton.}. This symmetry restricts the allowed operators in the effective action and protects the flatness of the potential, thereby maintaining the slow-roll conditions required for inflation. In natural inflation~\cite{Freese:1990rb} and axion monodromy models~\cite{Silverstein:2008sg}, the inflaton arises as a pseudo-Nambu–Goldstone boson with a softly broken shift symmetry, while in the Effective Field Theory of Inflation (EFToI)~\cite{Cheung:2007st}, the Goldstone mode associated with broken time translations inherits an approximate shift symmetry that constrains the structure of allowed interactions.

As a result, contributions from photons and gravitons to $\psi_n$ do not experience anomalous dimension running, and the phase rule derived from \CRT symmetry continues to apply non-perturbatively. This lends further support to the robustness of the no-go theorem in models composed solely of massless spin-1 and spin-2 fields. The status of scalar fields, including the inflaton, depends sensitively on their symmetry structure and must be assessed on a case-by-case basis.

\subsection{Cosmological Analogue of Furry's Theorem}\label{sec:CosmoFurrysTheorem}
An interesting observation is that an implication of the no-go theorem relates to a well-known result in QED known as Furry’s theorem, which states that any correlation function of $n \in 2\mathbb{Z}+1$ photon operators must vanish~\cite{Furry:1937zz}.

\subsubsection*{Review of Furry's Theorem in Flat Space}
Furry’s theorem in flat space asserts that any Feynman diagram containing a closed fermion loop with an odd number of vertices contributes zero to the total amplitude. This result arises if there is a charge conjugation symmetry, under which the charge conjugation operator $\mathbf{C}$ anticommutes with the photon field $A^{\mu}(x)$\footnote{Even though the photon field $A^{\mu}(x)$ can be expressed in terms of real components, it still transforms with a minus sign under charge conjugation \C. This transformation property is not determined by whether the field is real or complex in a given basis, but by how it couples to charged matter. \C is a discrete symmetry operation that inverts the sign of all charges. In particular, it sends particles to antiparticles and reverses the sign of their electromagnetic interactions. To preserve the structure of the theory under this transformation, the gauge field must also transform appropriately. For example, consider the QED interaction term between the photon and a Dirac fermion:
\begin{equation}
    \mathcal{L}_{\text{int}} = -e \bar{\psi} \gamma^\mu A_{\mu} \psi , .
\end{equation}
Under \C, the fermion field transforms to its charge-conjugate $\psi \to \psi^c$, which carries the opposite charge. To ensure that the interaction term remains invariant under \C, the photon field must flip sign:
\begin{equation}
    \mathbf{C} A^{\mu}(x) \mathbf{C}^\dagger = -A^{\mu}(x) , .
\end{equation}
This transformation rule holds regardless of the field basis because it reflects the physical requirement that a particle and its antiparticle couple with opposite sign to the electromagnetic field. Thus, the minus sign picked up by $A^{\mu}$ under charge conjugation is a universal feature of gauge theories with conserved charge, and plays a central role in results such as Furry’s theorem.}:
\begin{equation}
    \mathbf{C} A^{\mu}(x) \mathbf{C}^{\dagger} = -A^{\mu}(x) \, .
\end{equation}
Since the vacuum state $| \Omega\rangle$ is invariant under charge conjugation $\mathbf{C}|\Omega\rangle = |\Omega\rangle$, the correlation function of a single photon operator satisfies
\begin{equation}
    \langle \Omega | A^{\mu}(x) | \Omega \rangle = \langle \Omega | \mathbf{C}^{\dagger} \mathbf{C} A^{\mu}(x) \mathbf{C}^{\dagger} \mathbf{C} | \Omega \rangle = -\langle \Omega | A^{\mu}(x) | \Omega \rangle \implies \langle \Omega | A^{\mu}(x) | \Omega \rangle = 0 \, .
\end{equation}
Extending this reasoning, the correlation function for any odd number of photon operators also vanishes:
\begin{equation}\label{eqn:Furry}
    \langle \Omega | A^{\mu_1}(x_1) A^{\mu_2}(x_2) \dots A^{\mu_{2n+1}}(x_{2n+1}) | \Omega \rangle = 0 \, .
\end{equation}
This reflects the fact that such amplitudes change sign under charge conjugation, and therefore must vanish when evaluated in a \C-invariant vacuum. In QED, this implies that any process involving an odd number of photon fields and/or currents must vanish --- whether the photons are on-shell or off-shell. This significantly simplifies amplitude computations by eliminating entire classes of diagrams. Since Furry’s theorem holds non-perturbatively, it also applies order by order in perturbation theory. At leading order, for instance, any fermion loop with an odd number of photon vertices gives a vanishing contribution to the amplitude.

\subsubsection*{Cosmological Extension and Interpretation}
From \eqref{eqn:PhaseFormula} and \eqref{eqn:PhaseArgument}, we observe that in even $D = d+1$-spacetime dimensions, $\psi_n \in i\mathbb{R}$ when the number of external photon legs is odd. Consequently, any parity-even correlator must vanish, suggesting that for any theory which respects parity \P (such as standard QED), cosmological correlators with an odd number of photons (and no charged particles) also vanish. We thus find a cosmological analogue of Furry’s theorem!\footnote{The author thanks Aron Wall for extensive discussions regarding Furry's theorem.}

Moreover, this result is robust under renormalisation. Gauge invariance forbids the generation of a photon mass term, ensuring that the conformal dimension $\Delta$ of the boundary operator dual to the bulk gauge field remains protected. In particular, the dimension $\Delta=d-1$ for a conserved spin-1 current is exact and receives no anomalous corrections. Just as in flat space, the transversality of the photon self-energy enforces that its anomalous dimension vanishes to all orders, so the wavefunction renormalisation does not introduce any logarithmic scale dependence. As a result, the phase of any $\psi_n$ involving only photons is also protected under renormalisation, and the vanishing of odd-$n$ correlators in parity-invariant theories is a non-perturbatively stable feature. This underlines the universality of the cosmological analogue of Furry’s theorem across energy scales and interaction strengths.

Additionally, although a vanishing flat-space amplitude\footnote{This is referred to as the ``flat-space limit'' in the cosmology literature, where the residue of the total-energy pole for a Feynman-Witten diagram coming from particular interaction in cosmology, is the corresponding massless flat-space amplitude since the amplitude corresponds to the ultraviolet regime of the associated flat-space process, where the masses of the internal propagators are effectively zero. See~\cite{Cespedes:2025dnq}, where the authors introduce a novel massive flat-space limit, in which the internal masses in the corresponding flat-space Feynman graph remain finite.} does not necessarily imply that the corresponding cosmological correlator must vanish, \eqref{eqn:Furry} suggests that if the theory respects charge conjugation \C, then cosmological correlators with an odd number of photons (and no charged particles) also vanish. This serves as a complementary non-perturbative derivation of Furry’s theorem in a cosmological setting.

\subsection{Odd Spacetime Dimensions}
In odd $D=d+1$-spacetime dimensions, parity-even correlators at the $\eta=0^-$ boundary of inflation \emph{cannot} be generated at \emph{tree level or even-loop level} by models with
\begin{itemize}
    \item \emph{only} massless scalar fields and even-spin fields (e.g. gravitons) satisfying the Bunch-Davies vacuum (can also have even number of massless odd-spin fields, e.g. photons, or conformally coupled fields in external legs);
    \item locally \CRT-invariant Lagrangians;
    \item IR-finite and UV-finite boundary wavefunction coefficients $\psi_{n}$.
\end{itemize}
This result is particularly intriguing, as computing cosmological correlators in odd spacetime dimensions poses significant challenges due to the mode functions being expressed in terms of Hankel functions, $H_{\nu}(-k\eta)$ where $\nu\in\mathbb{Z}$. Although certain computations, such as the tree-level contribution to the two-point function, can be carried out in general $D=d+1$-spacetime dimensions for fields of arbitrary mass (see, e.g.,\cite{Thavanesan:2025csh}), it would be compelling to explore whether this result could also be established using alternative methods, such as the Wick rotation method employed in~\cite{Stefanyszyn:2023qov}.

Another noteworthy observation we can make is that in $D=d+1$-spacetime dimensions, $\psi_n \in i\mathbb{R}$ at tree-level (and any even-loop level) for the case of pure gravity or massless scalars (and any massless even spin-$s$ fields). However, normalisability imposes the condition that $\psi_n \in \mathbb{R}<0$. This leads to the intriguing conclusion that bulk loop effects are necessary for a consistent theory of massless scalars and gravitons.

More broadly, this work extends and generalises the previous no-go theorems of~\cite{Liu:2019fag,Cabass:2022rhr,Stefanyszyn:2023qov} by providing statements that hold in any spacetime dimension for any massive integer spin-$s$ field, as long as the conformal dimension of the field, $\Delta$, satisfies $\Delta \in \mathbb{R}$.

\section{Summary and Outlook}\label{sec:Summary}
In this paper, we have established a comprehensive framework for understanding the constraints on parity-violating signals in primordial cosmology, particularly by deriving a novel No-go Theorem for Cosmological Parity Violation. Specifically, we showed that in even $D=d+1$ spacetime dimensions, parity-odd correlators cannot arise under the following conditions: (1) the theory involves only massless scalar fields and even-spin fields (e.g., gravitons) and/or an even number of conformally coupled or massless odd-spin fields (e.g. photons) satisfying the Bunch-Davies vacuum; (2) the Lagrangian is locally \CRT-invariant; and (3) boundary wavefunction coefficients $\psi_n$ are IR-finite and UV-finite. When these conditions are met, any parity-odd correlator must arise solely from the factorised contributions of massive spinning fields, as shown explicitly in Sections~\ref{sec:CosmoCorrelators} and~\ref{sec:WFUConstraints} and discussed in much greater detail in~\cite{Stefanyszyn:2023qov,Stefanyszyn:2024msm}. Additionally, we identified loop-level corrections in UV-divergent cases as a distinct pathway for generating parity-odd signals, with a key contribution arising from the correction associated with logarithmic UV-divergences which can be seen most naturally in dimensional regularisation.

While these results reinforce the stringent constraints on parity violation in scale-invariant theories with standard cosmological vacua, they also highlight exciting opportunities to probe parity violation by relaxing certain assumptions. Several concrete scenarios in which these constraints can be bypassed were discussed in Section~\ref{sec:WFUConstraints}. We elaborate further on these below.

\subsection{Yes-go Examples}
\subsubsection*{Dynamical Chern-Simons and Axion Inflation}
Interactions that induce IR-divergent wavefunction coefficients $\psi_n$, such as those appearing in Chern-Simons and Axion inflation models, naturally lead to parity-odd correlators. This is due to the presence of $\log(-k\eta)$ terms in $\psi_n$, as discussed in Section~\ref{sec:WFUConstraints}. Furthermore, most calculations in these models rely on a series expansion of the mode functions in terms of the chemical potential $\mu$ (see e.g.~\cite{Creque-Sarbinowski:2023wmb}), which when truncated modifies the form of the discrete symmetry constraints \CRT, \RR and $\mathbf{D}$, since only wavefunction coefficients computed in the full theory satisfies the form presented in this paper. A similar phenomena was found in the case of resonant non-Gaussianities~\cite{DuasoPueyo:2023kyh} where the expansion in the dimensionless frequency of oscillations $\alpha=\omega/H$ requires the Cosmological Optical Theorem (COT)~\cite{COT,Goodhew:2021oqg,Cespedes:2020xqq} to be modified
\begin{equation}\label{eqn:ModifiedCOT}
    \text{Modified COT: } \qquad \psi_n (k,\alpha) + \left[ \psi_n (-k^*,-\alpha) \right]^* = 0 \, ,
\end{equation}
where $\alpha=\omega/H\in\mathbb{R}^+$, and in the resonance approximation, for which terms that are exponentially suppressed in $\alpha$ are neglected. For such an approximation \CRT and \RR would also be modified. In the case of the approximations used for theories with chemical potentials, where terms suppressed in $\mu$ are neglected, one would find \RR and \CRT to be modified in the following way\footnote{I would like to thank Harry Goodhew and Tommaso Moretti for helpful discussions regarding this point during the \href{https://parity.cosmodiscussion.com/}{Parity Violation from Home 2024} conference.}:
\begin{align}
    \text{Modified } \mathbf{RR} \qquad  \left[ \psi^{(L)}_n(\bfk;\mu) \right]^* &= e^{i\pi [(d+1)L-1]} \psi^{(L)}_n(e^{-i\pi}\bfk;e^{-i\pi}\mu) \, , \label{eqn:ModifiedRR} \\
    \text{Modified } \mathbf{CRT} \qquad  \left[\psi^{(L)}_n(\bfk;\mu)\right]^* &= e^{i\pi [(d+1)(L-1) + dn - \sum_\alpha \Delta_\alpha]} \psi^{(L)}_n(\bfk;e^{-i\pi}\mu) \, , \label{eqn:ModifiedCRT}
\end{align}
i.e. \CRT will no longer relate $\psi_n$ to its complex conjugate, and thus can no longer be used to determine the phase of  $\psi_n$. This implies that the discrete symmetry of \CRT involves an analytic continuation of the chemical potential, despite this not being the case for the original full theory. Hence, this suggests that the complex phase of $\psi_n$ and consequently the non-vanishing parity-odd correlators found in these theories are an artefact of the expansion since these models would fall within the class of models included in the regime of this paper's no-go theorem. Interestingly, \CRT symmetry of the original Lagrangian implies that for inflationary models involving a chemical potential, there must be a fundamental relation between the chemical potential $\mu$ parametrising $\mathbf{CP}$-violation and the slow-roll parameter $\xi$ which is a measure of $\mathbf{T}$-violation. Given that we already have bounds on the $\xi$ from the spectral tilt of the 2-point function, this could potentially lead to constraints on baryogenesis for these inflationary models. These all suggest further avenues to refine these analyses to study and probe parity-violating signatures.

\subsubsection*{Non-BD Vacuum States}
The discrete symmetries \CRT, \RR and $\mathbf{D^{\pm}_{-1}}$ defined in this paper and~\cite{Goodhew:2024eup} of the cosmological wavefunction hold strictly for the Bunch-Davies vacuum and $\alpha$-vacua with real Bogoliubov coefficients (see~\cite{Ghosh:2024aqd} where this was derived perturbatively). Relaxing this condition by allowing for non-BD initial states breaks \CRT symmetry, resulting in parity-odd signals in the final state correlators. This offers a compelling direction for future investigations into the role of initial conditions in generating parity violation.

\subsubsection*{Massive Spinning Fields}
Parity-odd correlators can emerge through factorised contributions when massive internal spinning fields are present. This phenomenon arises from the universal reality of wavefunction coefficients' total-energy poles under tree-level and UV-finite conditions. Such factorised contributions are a robust signal of parity violation and provide a clean observational target and are discussed in further detail in~\cite{Stefanyszyn:2023qov,Stefanyszyn:2024msm}.

\subsubsection*{Dimensional Regularisation for Loop Diagrams}
We have shown that UV-divergent $1$-loop diagrams can generate parity-odd signals via an $i\pi$ correction in dimensional regularisation, as previously noted in~\cite{Lee:2023jby}. However, we identify that this is subject to the choice of regularisation scheme one uses and highlight a seeming contradiction between the results obtained using the regularisation scheme of~\cite{Senatore:2009cf} and~\cite{Melville:2021lst,Lee:2023jby}. This outcome not only extends earlier calculations but also emphasises the necessity of harmonising various regularisation methods, particularly about how field masses are renormalised. A comprehensive analysis of these aspects will be provided in~\cite{Thavanesan:2025tha}.

\subsection*{Extensions to General FLRW Spacetimes}
The results derived in flat FLRW spacetimes extend naturally to more general cosmological backgrounds. The solutions to the $\eta \to 0$ limit of the equations of motion for general power-law cosmologies suggest that equivalent constraints on the wavefunction phase hold in broader settings. It would also be interesting to investigate the implications of \CRT for curved FLRW cosmologies given that the overall curvature of the universe is still an open question in cosmology (see e.g.~\cite{DiValentino:2019qzk,Handley:2019tkm,Handley:2019anl,Avis:2019eav,Thavanesan:2020lov,Shumaylov:2021qje,Letey:2022hdp,Huang:2022mxj,Bel:2022iuf,Ratra:2022ksb,Vigneron:2022bgr,Dineen:2023nbt,Vigneron:2024bfj} for recent studies on curved inflating universes). Investigating these generalisations will likely uncover new mechanisms for generating parity-violating signals.

\subsection{Future Directions}
This work serves as a foundation for both theoretical and observational advancements in probing parity violation in the early universe. Theoretically, future studies could explore non-standard inflationary scenarios, such as those with time-dependent couplings, ghost condensates (see e.g.~\cite{GhostCondensate,Arkani-Hamed:2003juy,Cabass:2022rhr,Cabass:2022oap}), or deviations from exact scale invariance, i.e. slow-roll corrections due to time translation invariance being slightly broken by standard inflation. Observationally, these insights could guide the analysis of forthcoming data from CMB polarisation and large-scale structure surveys, where parity-violating signals may manifest.

By clarifying the conditions under which parity-odd correlators arise and identifying concrete examples of their generation, this paper contributes to our understanding of the fundamental symmetries of the universe and opens pathways to uncover new physics beyond the standard models of cosmology and particle physics. The interplay between theoretical constraints and observational possibilities underscores the importance of continuing this line of inquiry, with the ultimate goal of unravelling the mysteries of the primordial universe.


\section*{Acknowledgements}
I am deeply grateful to Carlos Duaso Pueyo for his initial encouragement and confidence, which were instrumental in driving my research program on CPT in cosmology. I also thank Juan Maldacena for posing the initial question regarding the implications of CPT for chiral gravity waves. I am indebted to Cyril Creque-Sarbinowski, Harry Goodhew, Jiamin Hou, Lorenzo Di Pietro, Ciaran McCulloch, Tommaso Moretti, Oliver Philcox, Guilherme Pimentel, Marko Simonović, Zachary Slepian, David Stefanyszyn, Neeraj Tata, Marija Tomašević, Xi Tong, and Yuhang Zhu for their engaging discussions on parity violation. I am especially thankful to the organisers of the \href{https://parity.cosmodiscussion.com/}{Parity Violation from Home 2024} conference for the opportunity to present this work before its release and for the stimulating exchanges with participants that followed. I also appreciate the valuable feedback from Tarek Anous, Alejandra Castro, Vasudev Shyam, David Stefanyszyn, and Aron Wall.

I am grateful to New York University Abu Dhabi (NYUAD) for hosting the Strings 2025 conference, where this paper was finalised, and for their warm hospitality and financial support through the Strings 2025 fellowship. I am supported by the Bell Burnell Graduate Scholarship Fund, the Cavendish (University of Cambridge), the AFOSR grant FA9550-19-1-0260, “Tensor Networks and Holographic Spacetime”, the Heising-Simons Foundation, the Simons Foundation, and grants no. NSF PHY-2309135 to the Kavli Institute for Theoretical Physics (KITP), as well as a KITP Graduate fellowship. For the purpose of open access, I have applied a CC BY public copyright licence to any Author Accepted Manuscript version arising.

\bibliographystyle{JHEP}
\bibliography{refs}

\end{document}